\documentclass[twocolumn,fp]{jpsj3}

\usepackage{graphicx}
\usepackage{dcolumn}
\usepackage{bm}
\usepackage{color}
\usepackage{ulem}
\usepackage{mathrsfs}
\usepackage{txfonts}
\def\vector#1{{\boldsymbol{#1}}}

\def\vk{{\vector k}}

\def\vR{{\vector R}}

\def\TAF{{T_{\rm AF}}}

\def\hsp#1{\hspace{#1ex}}

\def\gsim{\stackrel{{\textstyle>}}{\raisebox{-.75ex}{$\sim$}}}

\def\eq.#1{Eq.~(\ref{#1})}
\def\eqs.#1{Eqs.~(\ref{#1})}

\def\Hc2{{H_{\rm c2}}}
\def\difHc2{{H'_{\rm c2}}}

\def\Neel{N\'{e}el}
\def\Neelpipi{N\'{e}el-$(\pi,\pi)$}
\def\Neelpizero{N\'{e}el-$(\pi,0)$}
\def\Neelzeropi{N\'{e}el-$(0,\pi)$}

\def\rimb{r_{\rm imb}}

\def\PK{${\rm P}_{\rm K}$}
\def\PM{${\rm P}_{\rm M}$}
\def\PKprim{${\rm P}_{\rm K}'$}
\def\PMprim{${\rm P}_{\rm M}'$}

\def\*ref*{{\color{red}$\leftarrow$Ref.{\bf [~~~~~]}}}

\newcommand\Equation[2]{
\begin{equation}\label{#1} 
#2
\end{equation}
}

\newcommand\Equationnoeqn[2]{
\begin{equation*}\label{#1} 
#2
\end{equation*}
}

\title{
Magnetic Structures of Electron Systems on the Extended Spatially Completely \\ 
Anisotropic Triangular Lattice near Quantum Critical Points \\ 
}

\author{Yuki Kono$^1$ 
and Hiroshi Shimahara$^2$\thanks{Corresponding author. 
E-mail: hiro@hiroshima-u.ac.jp}
}

\inst{
$^1$Graduate School of Advanced Sciences of Matter, Hiroshima University, \\ 
Higashi-Hiroshima, Hiroshima 739-8530, Japan \\ 
$^2$Graduate School of Advanced Science and Engineering, Hiroshima University, \\ 
Higashi-Hiroshima, Hiroshima 739-8530, Japan
}

\recdate{June 6, 2021}

\abst{
We examine magnetic structures of electron systems on 
an extended triangular lattice that consists of two types 
of bond triangles with electron transfer energies 
$t_{\ell}$ and $t_{\ell}'$ ($\ell = 1$, 2, and 3), respectively. 
We examine the ground state in the mean-field theory when $t_1 = t_1'$, 
focusing on collinear states with two sublattices. 
It is shown that 
when the imbalance of the spatial anisotropies 
of the two triangles is large, 
up-up-down-down (uudd) phases are stable, 
and 
the most likely ground states of 
the $\lambda$-${\rm (BETS)_2FeCl_4}$ system are 
the {\Neel} state with the modulation vector 
$(\pi/c,\pi/a)$ and a uudd state, 
where $c = a_1 = a_1'$ and $a = (a_2+a_2')/2$, 
with $a_1$, $a_1'$, $a_2$, and $a_2'$ being 
the lattice constants of the bonds with 
$t_1$, $t_1'$, $t_2$, and $t_2'$, respectively. 
These results are consistent with those from the classical spin system. 
In addition, this study reveals behaviors near 
the quantum critical point, which cannot be reproduced
in the localized spin model. 
As the imbalance of the spatial anisotropies increases, 
the $U_{\rm c}$ of the {\Neel} state increases, 
and that of the uudd state decreases. 
In the phase diagrams containing areas of 
the paramagnetic state, 
the {\Neel} state with $(\pi/c,\pi/a)$, 
and a uudd state, 
their boundaries terminate at a triple point, 
near which all the transitions are of the first order.
The phase boundary between the antiferromagnetic phases does
not depend on $U$, 
and the transition is of the first order everywhere 
on the boundary.
By contrast, 
the transitions from the two antiferromagnetic phases 
to the paramagnetic phase are of the second order, 
unless the system is close to the triple point.
}

\begin{document}
\sloppy
\maketitle

\section{\label{sec:introduction}
Introduction 
}

\begin{figure}[htbp]
\begin{center}
\begin{tabular}{cc}
{\small (a)} & 
\includegraphics[width=4.5cm]{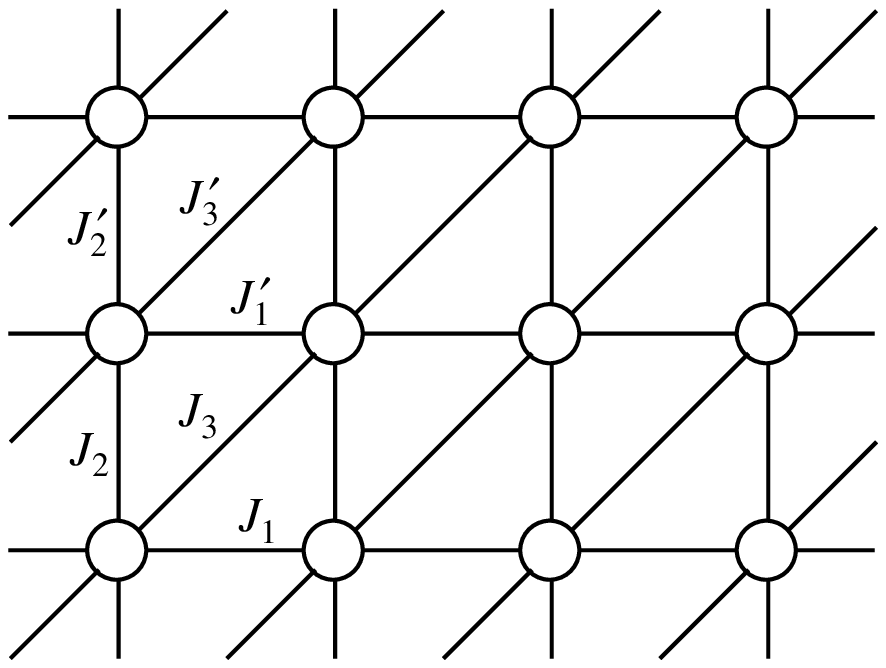}~~~
\\ 
\\ 
{\small (b)} & 
\includegraphics[width=4.5cm]{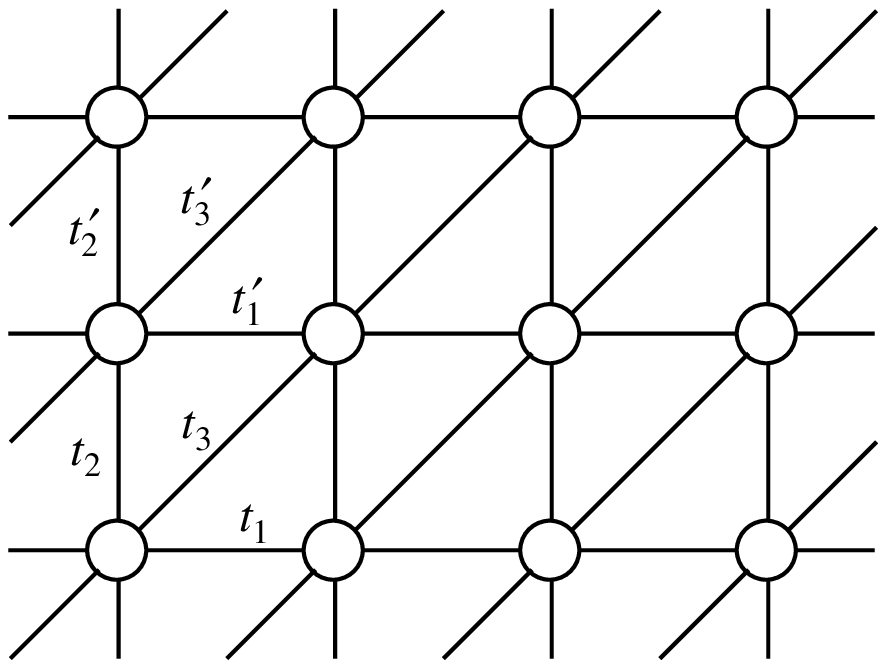}~~~
\end{tabular}
\end{center}
\caption{
Extended completely spatially anisotropic triangular lattice 
and definitions of 
(a) exchange coupling constants 
in localized spin systems 
and 
(b) transfer integrals 
in itinerant electron systems. 
} 
\label{fig:ESCATL}
\end{figure}

Electron systems on triangular lattices 
have been extensively studied because of intriguing phenomena, 
such as a possible quantum spin liquid, 
magnetic plateaus, 
spiral spin structures, 
and rich magnetic phase diagrams,~\cite{Bal10,Kan11} 
all of which originate from frustration in spin alignment. 
The frustration is maximum 
when all the antiferromagnetic interactions on the bonds are equal; 
hence, spatial anisotropy, or inequality, of the interactions 
reduces the frustration. 
However, the influence of the spatial anisotropy 
can be practically significant when real compounds are examined.~\cite{Hau13} 
For example, some organic compounds contain 
spatially anisotropic triangular lattices, shown in Fig.~\ref{fig:ESCATL}, 
that consist of two types of bond triangles; 
these lattices are called 
extended spatially completely anisotropic triangular lattices (ESCATLs). 
The model for localized spin systems has six types of exchange interactions 
with coupling constants $J_{\ell}$ and $J_{\ell}'$, 
as shown in Fig.~\ref{fig:ESCATL}(a), 
while the model for itinerant electron systems has six types 
of transfer integrals $t_{\ell}$ and $t_{\ell}'$, 
as shown in Fig.~\ref{fig:ESCATL}(b), where $\ell = 1$, 2, and 3.

Interestingly, the ESCATL includes some frustrated lattices 
as special cases. 
It reduces to the spatially completely anisotropic triangular lattice (SCATL) 
when $J_{\ell} = J_{\ell}'$ for all $\ell$,~\cite{Hau13} 
and spatially anisotropic triangular lattice (SATL) 
when $J_{\ell} = J_{\ell}'$ for all $\ell$ 
and $J_2 = J_3$.~\cite{Kan11} 
When $J_3' = 0$, the ESCATL reduces to the trellis lattice,~\cite{Gop94} 
and when $J_3' = J_1 = J_1' = 0$, 
it reduces to the honeycomb lattice.~\cite{Cas09} 
Compared with these reduced lattices, 
the ESCATL has a unique feature 
of an imbalance in the spatial anisotropies 
of the two bond triangles.~\cite{Sak17}

\begin{figure}[htbp]
\begin{center}
\begin{tabular}{c}
\includegraphics[width=6.0cm]{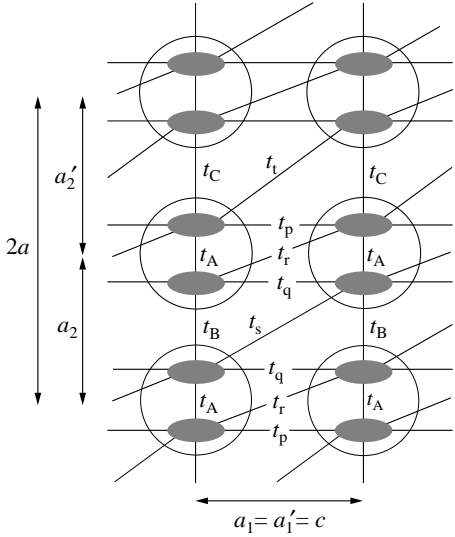} 
\end{tabular}
\end{center}
\caption{
Schematic of the crystal structure of the BETS layers 
in the $\lambda$-$X$ system 
and the ESCATL. 
The small, dark ellipses 
and large circles represent 
BETS molecules and their dimers, respectively. 
The hopping integrals 
$t_{\rm A}$, 
$t_{\rm B}$, 
$t_{\rm C}$, 
$t_{\rm p}$, 
$t_{\rm q}$, 
$t_{\rm r}$, 
$t_{\rm s}$, and 
$t_{\rm t}$ 
are as defined by Kobayashi and Mori.~\cite{NoteHot00} 
In terms of their hopping integrals, 
the hopping integrals in the present model are expressed 
as 
$t_1 = t_1' = (- t_{\rm p} - t_{\rm q} + t_{\rm r})/2$, 
$t_2 = t_{\rm B}/2$, 
$t_2' = t_{\rm C}/2$, 
$t_3 = t_{\rm s}/2$, and $t_3' = t_{\rm t}/2$. 
We refer to the lattice constants of the bonds with the hopping integrals 
$t_1$, $t_1'$, $t_2$, and $t_2'$ 
as 
$a_1$, $a_1'$, $a_2$, and $a_2'$, 
respectively, 
and define $c = a_1 = a_1'$ and $a = (a_2 + a_2')/2$. 
The lattice constant in the crystal a-axis of the compound 
corresponds to $2a$ in the present model. 
}
\label{fig:lambdaFe_lattice}
\end{figure}

\begin{figure}[htbp]
\begin{center}
\begin{tabular}{cc}
\multicolumn{2}{c}{
  \begin{tabular}{c}
\includegraphics[width=3.6cm]{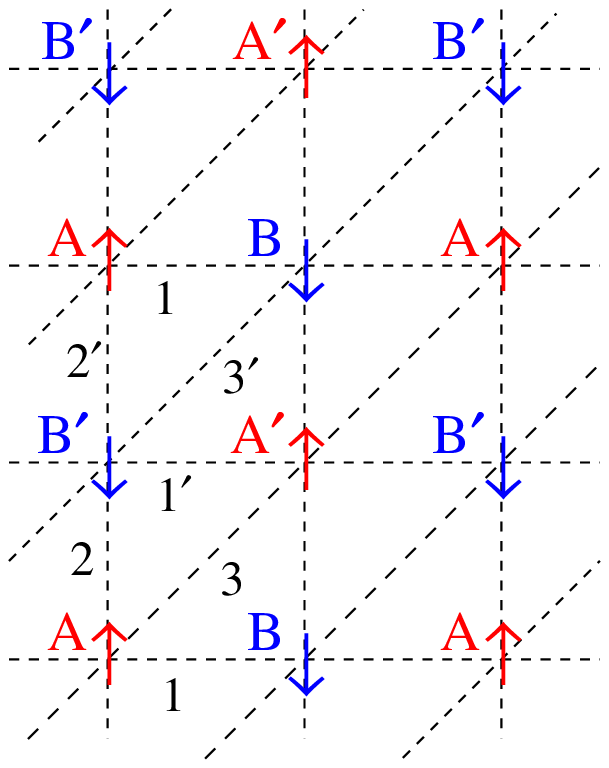} \\ 
      {\footnotesize (a) {\Neelpipi} state}
      \\[12pt] 
  \end{tabular}
  } 
\\ 
  \begin{tabular}{c}
\includegraphics[width=3.6cm]{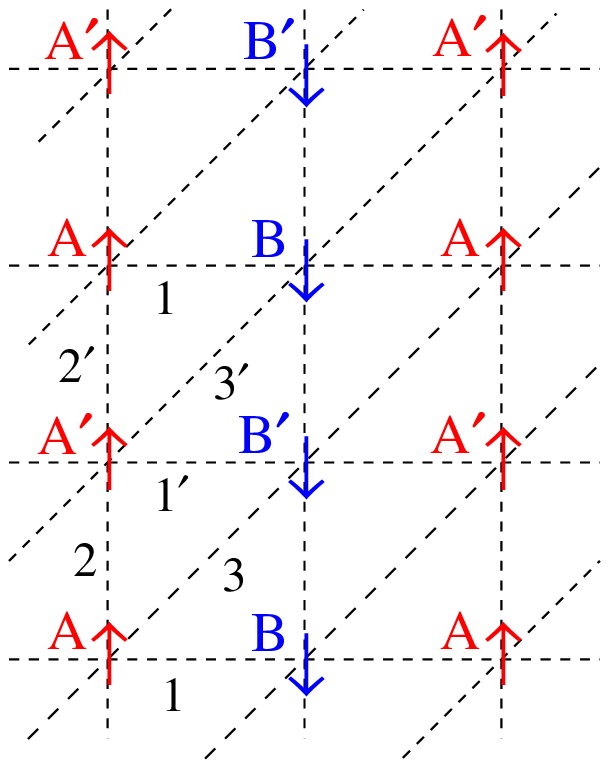} \\ 
      {\footnotesize (b) {\Neelpizero} state}
  \end{tabular}
  \hspace{-2ex} 
  & 
  \hspace{-2ex} 
  \begin{tabular}{c}
\includegraphics[width=3.6cm]{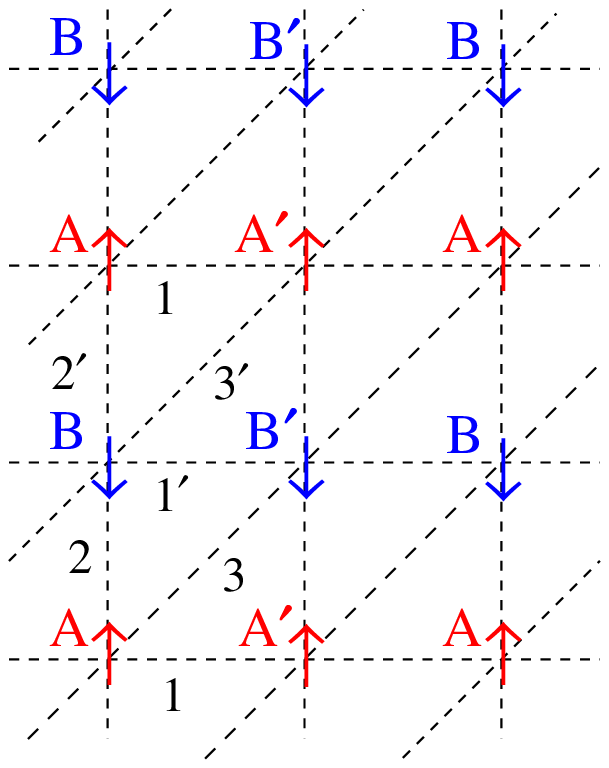} \\ 
      {\footnotesize (c) {\Neelzeropi} state} 
  \end{tabular}
\end{tabular}
\end{center}
\caption{
(Color online) 
Three {\Neel} states 
and the definitions of the sublattices. 
}
\label{fig:Neel}
\end{figure}

\begin{figure}[htbp]
\begin{center}
\begin{tabular}{cc}
\begin{tabular}{c}
\includegraphics[width=3.6cm]
{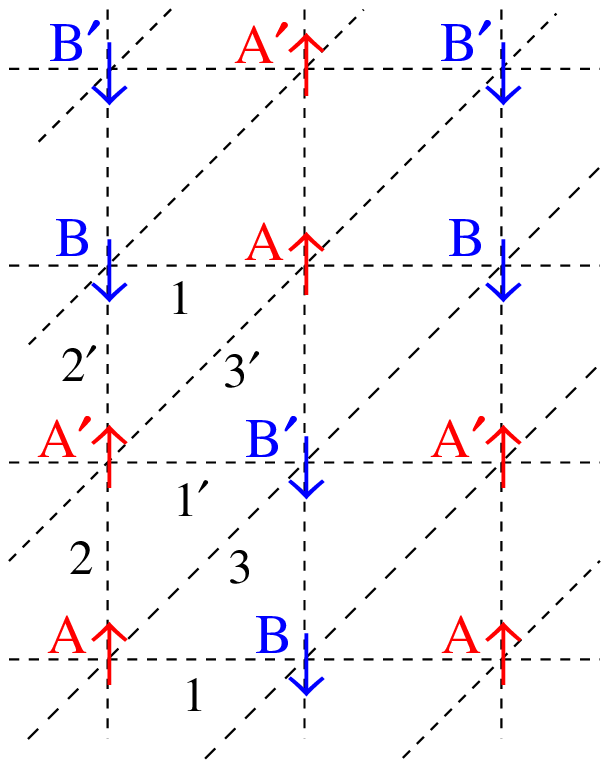} \\ 
{\footnotesize (a) uudd-2 state}
\end{tabular}
  \hspace{-2ex} 
  & 
  \hspace{-2ex} 
\begin{tabular}{c}
\includegraphics[width=3.6cm]
{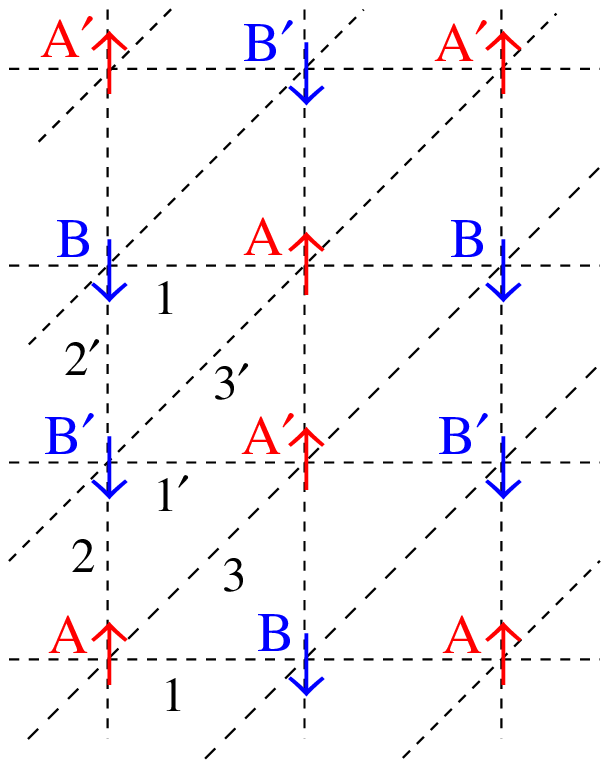} \\ 
{\footnotesize (b) uudd-$2'$ state}
\end{tabular}
\end{tabular}
\end{center}
\caption{
(Color online) 
Two uudd states and 
the definitions of the sublattices. 
} 
\label{fig:uudd} 
\end{figure}

In this study, 
we examine the itinerant electron systems on the ESCATL, 
being motivated by the organic compound 
$\lambda$-${\rm (BETS)_2}X{\rm Cl}_4$ 
(hereinafter abbreviated as $\lambda$-$X$), 
where $X = {\rm Fe}$, ${\rm Ga}$, 
and ${\rm Fe}_x{\rm Ga}_{1-x}$.~\cite{NoteBETS}  
These compounds have the ESCATL, 
as shown in Fig.~\ref{fig:lambdaFe_lattice}, 
if each dimer of the BETS molecules is regarded as a site. 
The purposes of the study are 
to extend our knowledge on the ESCATL antiferromagnet and 
to gain theoretical insight into the magnetic structure 
in the $\lambda$-$X$ systems. 
In the $\lambda$-$X$ systems 
that exhibit an antiferromagnetic long-range order, 
it is considered that itinerant 
$\pi$-electrons principally sustain the order, 
and localized d-spins are passive in the exchange fields 
created by the $\pi$-electrons~\cite{Notepid}, 
which is the reason why we examine the itinerant electron model. 
We consider a realistic situation 
in which the spiral spin structure is suppressed; 
hence, we focus on collinear antiferromagnetic spin structures, 
such as the three {\Neel} states defined in Fig.~\ref{fig:Neel} 
and the two up-up-down-down (uudd) states defined in Fig.~\ref{fig:uudd}. 
We examine the ground state in the mean-field theory. 
To parameterize the imbalance of the spatial anisotropies mentioned above, 
we adopt 
\Equationnoeqn{eq:rimb}
{
     \rimb \equiv \frac{t_3/t_2 - t_3'/t_2'}{t_3/t_2} 
     } 
because $t_1 = t_1'$ is satisfied in the $\lambda$-Fe system.~\cite{Mor02}

Systems on the ESCATL are realized in other organic compounds 
such as $\lambda$-${\rm (BEST)_2}X{\rm Cl}_4$ 
and $\lambda$-${\rm (BEDT}$-${\rm STF)_2}X{\rm Cl}_4$ 
as well as the $\lambda$-$X$ systems, 
when each dimer of the molecules 
is regarded as 
a single lattice site.~\cite{NoteSTF,NoteBEST,Mor02} 
In organic compounds of the form $D_2A$, where $D$ and $A$ represent 
a donor and an anion, respectively, 
the ESCATL can be realized 
when the donors are dimerized and 
the anions are in a staggered order. 
We examine wide parameter ranges, 
considering the potential for undiscovered compounds; however, 
we include candidate parameter sets 
for the $\lambda$-Fe system in the ranges 
so that the resulting phase diagrams include the points 
for this compound.

For the localized spin system on the ESCATL, 
classical phase diagrams~\cite{Sak17} contain areas of 
the five collinear antiferromagnetic phases defined 
in Figs.~\ref{fig:Neel} and \ref{fig:uudd} 
and the spiral spin phase. 
It was shown that the imbalance of the anisotropic parameters 
$J_2/J_3$ and $J_2'/J_3'$ stabilizes the uudd phases. 
Although similar uudd phases emerge in other systems, 
they are induced by 
four-spin,~\cite{Rog80} 
ferromagnetic,~\cite{Kim03,Mun01} 
and biquadratic exchange interactions.~\cite{Kap09,Zou16} 
The present uudd phases are unique as 
they are induced solely by antiferromagnetic exchange interactions, 
where the other types of interactions 
are not required. 
Below, we clarify whether the uudd phases of this mechanism occur 
in itinerant electron systems as well as 
in the localized spin system.

{\it Application to the $\lambda$-Fe system} --- 
In the next several paragraphs, 
we explain the background related to 
the $\lambda$-Fe system. 
The bonds along the crystal c- and a-axes of this system 
correspond to 
the $1$- and $1'$-bonds and $2$- and $2'$-bonds, 
respectively, 
as shown in Figs.~\ref{fig:ESCATL}--\ref{fig:uudd}. 
Although they are depicted as perpendicular for convenience, 
the c- and a-axes are not perpendicular 
in the $\lambda$-Fe system. 
The shear distortion gives rise to 
nonzero $t_3$ and $t_3'$ in the present model.

In the classical spin model with parameter values 
for the $\lambda$-Fe system,~\cite{Mor02} 
a spiral spin state has the lowest energy;~\cite{Sak17} 
however, it would be reasonable to assume that 
this state is suppressed in this system 
because experimental studies suggest that the ground state is 
a collinear antiferromagnetic state.~\cite{Osh17,Sas01,Tok05,Aki11,Sat98,Tok97,Ito16} 
In particular, an electron spin resonance (ESR) study suggests the existence 
of two sublattices in the magnetic field.~\cite{Osh17} 
The possible reasons for the suppression of the spiral spin state 
are factors not incorporated in the classical spin model, 
such as quantum fluctuations and 
anisotropy in the exchange fields created by the d-spins. 
When the spiral spin state is excluded, 
the lowest-energy state is the \Neel-$(\pi,\pi)$ state, 
and interestingly, 
the uudd-2 state has the second-lowest energy, 
which is slightly larger than the lowest energy. 
Because their difference is small and 
the estimated values of the energies contain errors due to 
simplifications in the model 
and errors in assumed parameters, 
it is reasonable to regard 
the \Neel-$(\pi,\pi)$ and uudd-2 states 
as candidates for the ground state of the $\lambda$-Fe system.

\begin{figure}[htbp]
\begin{center}
\begin{tabular}{cc}
\includegraphics[width=3.8cm]
{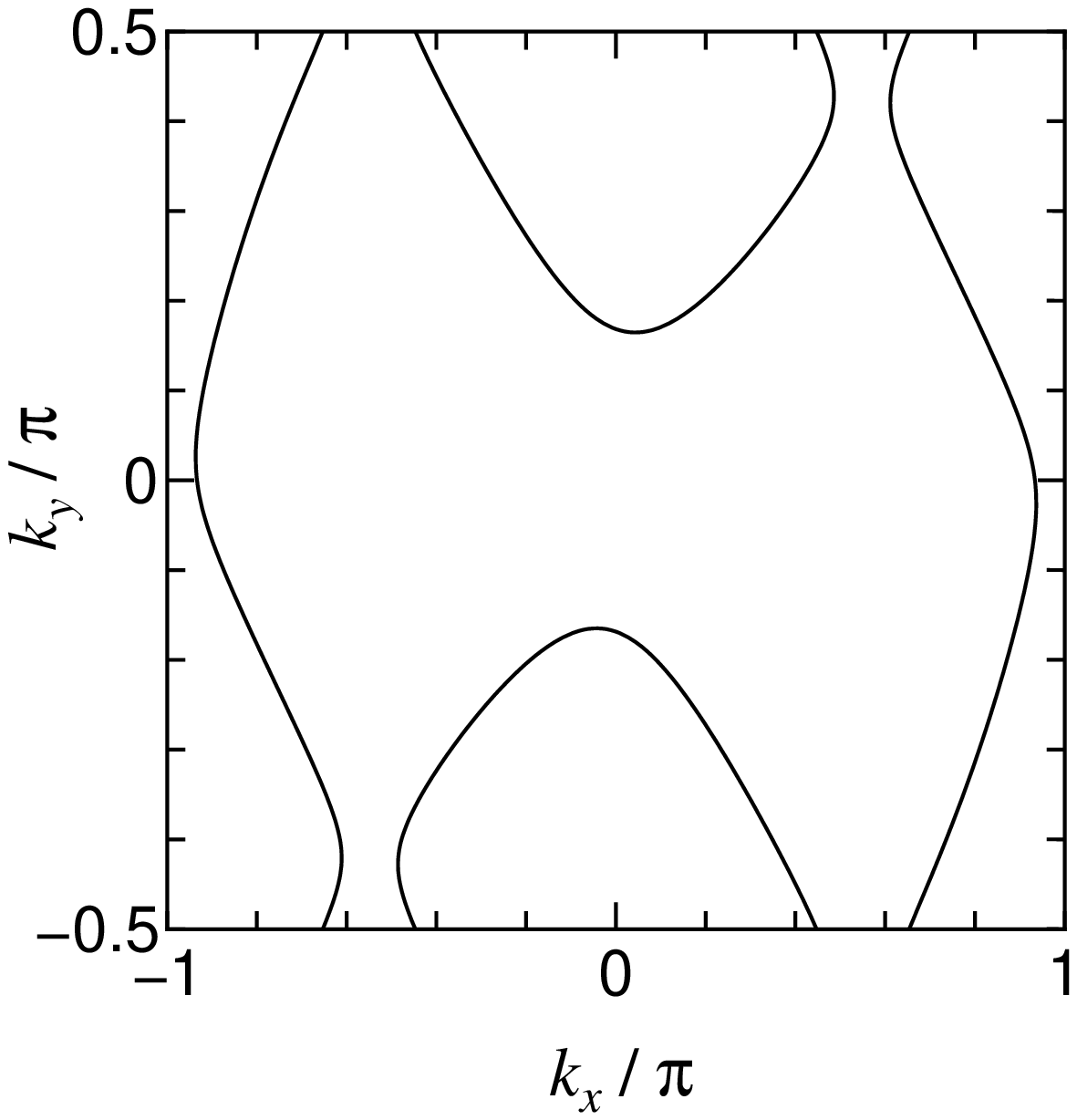} 
& 
\includegraphics[width=3.8cm]
{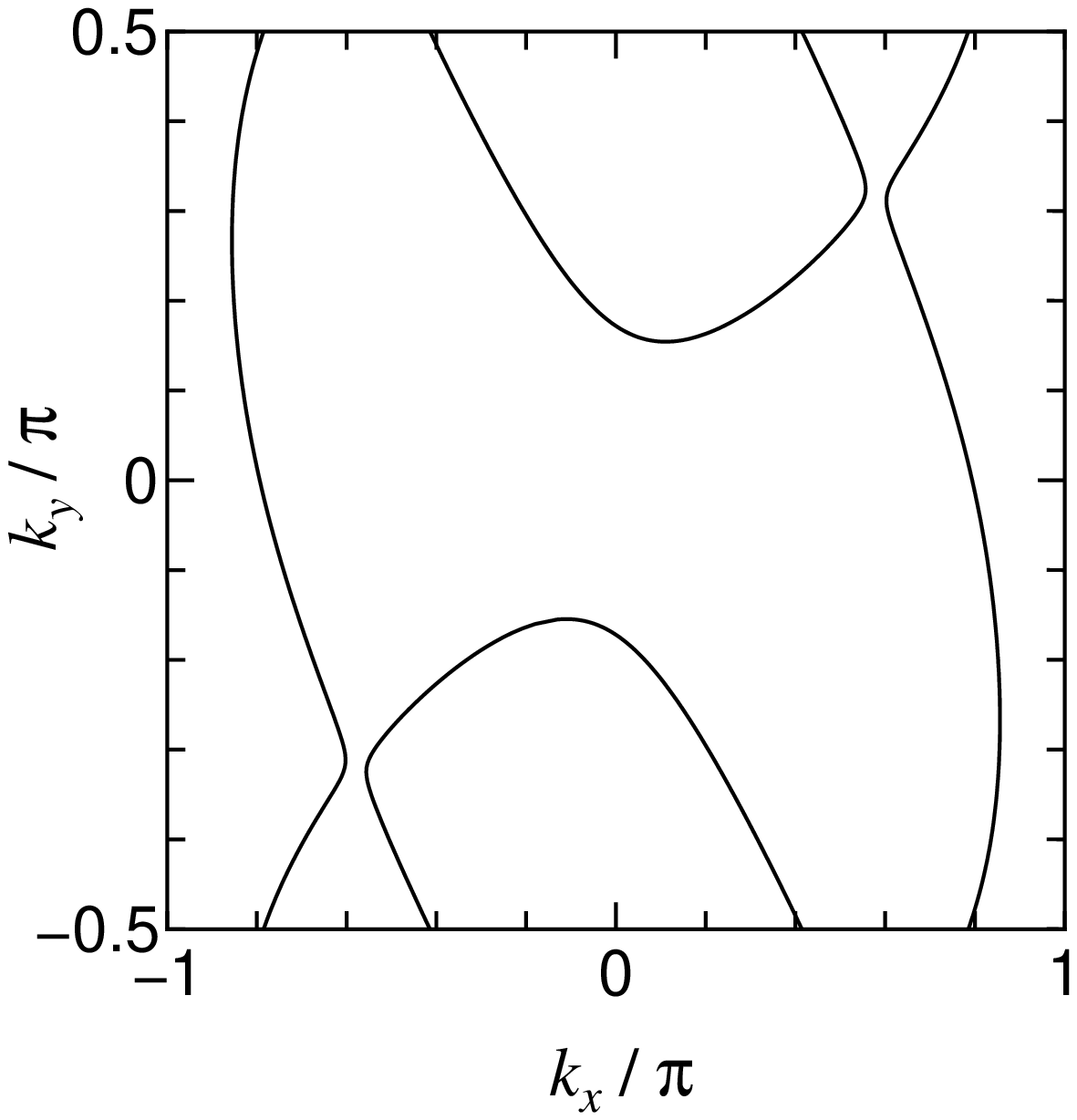} 
\end{tabular}
\end{center}
\caption{
Fermi surfaces for parameter sets {\PK} (left panel) 
and {\PM} (right panel) 
defined in Table~\ref{table:PMPK}. 
} 
\label{fig:FS} 
\end{figure}

\begin{table}[hbtp]
\caption{
Values of transfer integrals in units of $10^{-2}$~eV. 
Parameter sets {\PK} and {\PM} 
are based on the values of transfer integrals 
obtained by Kobayashi and Mori, respectively.~\cite{NoteHot00} 
The expressions for the transfer integrals $t_{\ell}$ and $t_{\ell}'$ 
are given in the caption 
of Fig.~\ref{fig:lambdaFe_lattice}. 
Parameter sets {\PKprim} and {\PMprim} are the variations of 
{\PK} and {\PM} in which $t_3$ is a variable 
and the other transfer integrals remain unchanged. 
}
\label{table:PMPK} 
{\small 
\begin{center}
\begin{tabular}{c}
\\ 
\begin{tabular}{c|p{1.2cm}p{1.2cm}p{1.2cm}p{1.2cm}}
\hline 
\hspace{-2ex}
\begin{tabular}{c}
Transfer \\[-2pt]
integral 
\end{tabular}
\hspace{-2ex}
       & \multicolumn{1}{c}{\PK}  
       & \multicolumn{1}{c}{\PKprim}  
       & \multicolumn{1}{c}{\PM}  
       & \multicolumn{1}{c}{\PMprim}  
       \\ 
\hline 
$t_1$  & ~4.6325         & ~4.6325      &  ~6.295          &  ~6.295    \\ 
$t_2$  & ~5.7555         & ~5.7555      &  ~5.29           &  ~5.29     \\ 
$t_3$  & ~2.535          &  Variable    & ~5.965           &  Variable  \\
$t_2'$ &  ~4.145         &  ~4.145      & ~6.205           & ~6.205     \\ 
$t_3'$ &  ~0.1955        &  ~0.1955     & ~0.965           & ~0.965     \\ 
\hline 
$\rimb$  &  ~0.891 & $\rimb(t_3)$   & ~0.862    &  $\rimb(t_3)$  \\ 
\hline 
\end{tabular}
\end{tabular}
\end{center}
}
\end{table}

The localized spin model can be an effective model for 
the $\lambda$-Fe system in the insulating phase; however, 
it misses the itinerant character of the $\pi$-electrons. 
In particular, they cannot take account of 
the fact that the $\lambda$-Fe system is possibly close to 
the quantum critical point ($U \approx U_{\rm c}$).~\cite{Shi17,NoteQCP} 
Here, 
$U$ and $U_{\rm c}$ are the on-site Coulomb energy 
and the critical value of $U$ 
between the antiferromagnetic and paramagnetic states, respectively. 
Hence, the itinerant electron system is worth studying 
for the $\lambda$-Fe system.

The magnetic structure of the $\lambda$-Fe system 
was studied by some authors using models in which 
the $\pi$-electrons are itinerant.~\cite{Bro98,Hot00,Ter04} 
Brossard et al. examined the magnetic structure 
of the d-spins in magnetic fields 
while treating the interaction between the d-spins 
in terms of the Ruderman-Kittel-Kasuya-Yosida (RKKY) interaction 
via conductive $\pi$-electrons.~\cite{Bro98} 
Terao and Ohashi examined the superconductivity and antiferromagnetism 
by considering the RKKY interaction.~\cite{Ter04} 
Hotta and Fukuyama examined the effect of d-spins on 
the $\pi$-electron system, 
considering the magnetic structure inside the dimers 
and obtained unified phase diagrams for some organic compounds~\cite{Hot00}. 
These studies were based on a physical picture in which 
the antiferromagnetic transition is induced principally in the d-spin system 
because they were conducted before later experimental studies 
denied this picture.~\cite{Aki09} 
This study is based on the current knowledge that 
the $\pi$-electron system is principal, whereas the d-spins are passive. 
Meanwhile, 
the magnetic structure inside the dimers~\cite{Hot00} 
is beyond the scope of this study.

Because the $\lambda$-Fe system has Fermi surfaces 
open in the $k_y$-direction as shown in Fig.~\ref{fig:FS}, 
it has often been suggested that 
the Fermi-surface nesting with the nesting vector 
near $(\pi/c,0)$ 
favors the antiferromagnetic order of this wave vector.\cite{Bro98,Mor02,Ter04} 
Here, it should be noted that the Fermi surfaces in Fig.~\ref{fig:FS} 
are drawn in the Brillouin zone halved by the difference between 
$(t_2,t_3)$ and $(t_2',t_3')$. 
The modulation vector $(\pi/c,0)$ in the half Brillouin zone 
cannot resolve the modulation vectors $(\pi/c,0)$ and $(\pi/c,\pi/a)$ 
in the original Brillouin zone, 
i.e., the {\Neelpizero} and {\Neelpipi} states 
defined in Figs.~\ref{fig:Neel}(a) and (b). 
Similarly, the modulation vector $(0,0)$ 
in the half Brillouin zone cannot resolve 
the modulation vectors $(0,0)$ and $(0,\pi/a)$ 
in the original Brillouin zone, 
i.e., 
the ferromagnetic state and 
the {\Neelzeropi} state defined in Fig.~\ref{fig:Neel}(c). 
The uudd phases have the modulation vector $(\pi/c,\pi/2a)$.

The experimental~\cite{Sat98,Kob99,Sat00,Aki12} 
and theoretical~\cite{Shi17} studies for 
the $\lambda$-${\rm Fe}_x{\rm Ga}_{1-x}$ system 
indicate that the influence of the d-spins is indispensable for 
the transition to 
the antiferromagnetic long-range order. 
This may appear inconsistent with the fact that the d-spins are passive; 
however, it can be explained on the basis of 
a stabilization effect~\cite{Shi14,Shi18} 
by the anisotropy in the spin space and/or the enhanced three dimensionality 
introduced by the d-spins. 
These factors are not {\it explicitly} incorporated 
in the present model for the pure $\pi$-electron system; however, 
the stabilization of the long-range order by these factors 
is {\it implicitly} assumed in the present mean-field approximation.

In Sect.~\ref{sec:formulation}, 
we present the model and formulation. 
In Sect.~\ref{sec:results}, 
we present the numerical results including phase diagrams. 
Section~\ref{sec:summary and discussion} 
summarizes and discusses the results.

\section{\label{sec:formulation}
Model and Formulation 
}

In the tight-binding model of the electron system on the ESCATL, 
electron energy dispersion can be written in the form~\cite{Hot00} 
\Equation{eq:epsilon}
{
     \epsilon_{\vk} 
     = 2 t_1 \cos k_x \pm \eta_{\vk}
     }
with 
\Equationnoeqn{eq:epsilon23} 
{
     \begin{split} 
     \eta_{\vk} 
     \equiv & \, \, 
         \Bigl \{  
                   [ (t_2 + t_2') \cos k_y + (t_3 + t_3') \cos (k_x + k_y)]^2 
                   \\[-8pt] 
            &
              ~~~    + [ (t_2 - t_2') \sin k_y + (t_3 - t_3') \sin (k_x + k_y)]^2 
              \Bigr \}^{1/2} . 
     \end{split}
     }
The lattice constants $a_1 = a_1'$ and $(a_2 + a_2')/2$ 
are absorbed into the definitions of 
the momentum components $k_x$ and $k_y$.

To obtain explicit results, 
we need explicit parameter values. 
Our plan for choosing parameter values is presented in 
Table~\ref{table:PMPK}. 
First, we adopt the sets of electron transfer energies 
{\PK} and {\PM} obtained by Mori and Kobayashi~\cite{NoteHot00}. 
Next, we extend the parameter region by varying $t_3$ 
from the values in {\PK} and {\PM}, 
keeping the other parameters fixed. 
We refer to the parameter sets in which $t_3$ is a variable as 
{\PKprim} and {\PMprim}. 
The shift in $t_3$ results in a shift in ${\rimb}$. 
Although we can shift ${\rimb}$ 
by shifting other hopping integrals, 
simply for the sake of convenience,
we shift only $t_3$. 
In the present dimer model, 
the $\pi$-electron band having Fermi surfaces is half-filled. 
The Fermi surfaces for these parameter sets {\PK} and {\PM} are depicted 
in Fig.~\ref{fig:FS}.

We consider the Hubbard Hamiltonian defined by 
\Equationnoeqn{eq:H}
{
     H = H_0 + H_1 
     }
with 
\Equation{eq:H0}
{
     H_0 = \sum_{i,j,\sigma}t_{ij} c_{i\sigma}^{\dagger} c_{j \sigma} 
             - \mu 
               \sum_{i}
               \left ( 
                 \sum_{\sigma} c_{i\sigma}^{\dagger} c_{i \sigma} - n 
               \right ) , 
     }
\Equationnoeqn{eq:H1}
{
     H_1 = U \sum_{i} {\hat n}_{i\uparrow} {\hat n}_{i \downarrow} , 
     }
where $c_{i\sigma}$ 
and $n$ are 
the annihilation operator of the electron on site $i$ with spin $\sigma$ 
and 
the number of electrons per site, 
respectively, 
and ${\hat n}_{i \sigma} = c_{i\sigma}^{\dagger} c_{i \sigma}$. 
The transfer energies $t_{ij}$ yield $\epsilon_{\vk}$ in \eq.{eq:epsilon}. 
The term $- \mu \sum_{i} (-n)$ in \eq.{eq:H0} is a constant in the sense that 
it does not contain any operators; however, 
it is relevant for deriving the correct self-consistent equation for $n$ 
[\eq.{eq:nSCE}]. 
Hence, we must retain the term in the evaluation of 
the total energy $E$.~\cite{Notemun}

We divide the lattice into four sublattices: 
A, B, ${\rm A}'$, and ${\rm B}'$, 
as shown in Figs.~\ref{fig:Neel} and \ref{fig:uudd}, 
which are defined so that the sites with the up-spins and down-spins 
belong to different sublattices. 
Moreover, 
we define additional sublattices 
that take into account the double periodicity 
due to $(t_2,t_3) \ne (t_2',t_3')$. 
For the {\Neelzeropi} state shown in Fig.~\ref{fig:Neel}(c), 
the additional sublattices ${\rm A}'$ and ${\rm B}'$ 
are not necessary; however, 
we defined them to unify the formalism. 
For all the states, 
the sublattices $X$ and $X'$ are eventually equivalent 
because of the spatial inversion; 
hence, 
unless spontaneous spatial-inversion-symmetry breaking occurs, 
the resulting states are two-sublattice states. 
Hence, the sublattice magnetization $m$ is defined by 
$m = m_{\rm A} = m_{\rm A'} = - m_{\rm B} = - m_{\rm B'}$ 
with 
\Equationnoeqn{eq:mdef}
{
     m_{X} \equiv \frac{1}{2} \left \langle {\hat n}_{i\uparrow} 
                                    - {\hat n}_{i \downarrow} \right \rangle 
     }
for $i \in X$. 
Because we consider neither the charge-density wave nor the charge order, 
$ \left \langle {\hat n}_{i\uparrow} 
+ {\hat n}_{i \downarrow} \right \rangle \equiv n$ 
is a constant independent of $i$; hence, 
$n_{i \sigma} 
 \equiv \left \langle {\hat n}_{i \sigma} \right \rangle 
 = n/2 + s_{\sigma} s_{X} m$ 
for $i \in X$, 
where 
$s_{\rm A} = s_{\rm A'} = 1$, $s_{\rm B} = s_{\rm B'} = -1$, 
$s_{\uparrow} = 1$, and $s_{\downarrow} = -1$. 
We define $c_{i\sigma}^{(X)} \equiv c_{i\sigma}$ for $i \in X$, 
where $X = {\rm A}$, ${\rm A}'$, ${\rm B}$, and ${\rm B}'$, respectively. 
The Fourier transformations are defined by 
\Equationnoeqn{eq:ckci}
{
     \begin{split}
     c_{\vk \sigma}^{(X)} = \sqrt{\frac{4}{N}} 
       {\sum_{i \in X}} e^{- i \vk \cdot \vR_i} 
     c_{i \sigma}^{(X)} , 
     \\ 
     c_{i \sigma}^{(X)} = \sqrt{\frac{4}{N}} 
       {\sum_{\vk}}' e^{i \vk \cdot \vR_i} 
     c_{\vk \sigma}^{(X)} , 
     \end{split}
     }
where $N$ denotes the number of the sites 
and the summation $\sum_{\vk}'$ is taken over 
a reduced Brillouin zone containing $N/4$ momenta 
(in the application to the $\lambda$-Fe system, 
$N$ is the number of the dimerized BETS sites).

The mean-field approximation 
\Equationnoeqn{eq:HUMF}
{
     H_1 = U \sum_{i} 
     \Bigl [ 
       \left \langle {\hat n}_{i\uparrow} \right \rangle {\hat n}_{i \downarrow} 
     + \left \langle {\hat n}_{i\downarrow} \right \rangle {\hat n}_{i \uparrow} 
     - \left \langle {\hat n}_{i\uparrow} \right \rangle 
       \left \langle {\hat n}_{i\downarrow} \right \rangle 
     \Bigr ] 
     }
leads to 
\Equationnoeqn{eq:HMFsubl}
{
     H = 
     {\sum_{\vk, \sigma}}' 
         \left ( \hsp{-0.8}
         \!\! 
         \begin{array}{cccc}
           c_{\vk \sigma}^{\rm (A) \dagger}, 
         \!\!\!&\!\!\!
           c_{\vk \sigma}^{\rm (A') \dagger}, 
         \!\!\!&\!\!\!
           c_{\vk \sigma}^{\rm (B) \dagger}, 
         \!\!\!&\!\!\!
           c_{\vk \sigma}^{\rm (B') \dagger} 
         \end{array}
         \!\! 
         \hsp{-0.5} 
         \right ) 
         {\hat {\cal E}}_{\vk \sigma} 
         \left ( \hsp{-0.5}
         \begin{array}{c}
         c_{\vk \sigma}^{\rm(A)} \\[4pt]
         c_{\vk \sigma}^{\rm (A')} \\[4pt]
         c_{\vk \sigma}^{\rm (B)} \\[4pt] 
         c_{\vk \sigma}^{\rm (B')} 
         \end{array} 
         \hsp{-0.8} 
         \right ) 
     + {\hat E}_0
     }
with 
\Equationnoeqn{eq:E0}
{
     {\hat E}_0(m,\mu) = - N U (\frac{n^2}{4} - m^2) + N \mu n . 
     } 
The elements ${\tilde \xi}_{\vk \sigma}^{(XY)}$ of 
${\hat {\cal E}}_{\vk \sigma}$ 
are shown in the Appendix\ref{app:ExplicitE}. 
With an appropriate unitary matrix, 
${\hat {\cal E}}_{\vk \sigma}$ is diagonalized as 
\Equationnoeqn{eq:diagU}
{
     E_{\vk \sigma}^{(\nu)} 
     = \sum_{X_1,X_2} 
       [u_{\vk \sigma}^{(X_1 \nu)}]^{*} 
         {\tilde \xi}_{\vk \sigma}^{(X_1X_2)} 
        u_{\vk \sigma}^{(X_2 \nu)} , 
     }
where $u_{\vk \sigma}^{(X\nu)}$ are the matrix elements 
of the unitary matrix 
and $E_{\vk \sigma}^{(\nu)}$ are the eigenvalues. 
Hence, the Hamiltonian is diagonalized as 
\Equationnoeqn{eq:diagH}
{
     \begin{split}
     H = 
         & 
           \sum_{\nu} 
           {\sum_{\vk, \sigma}}' 
             E_{\vk \sigma}^{(\nu)} \gamma_{\vk \sigma}^{(\nu) \dagger} 
                                    \gamma_{\vk \sigma}^{(\nu)}
           + {\hat E}_0(m, \mu) , 
     \end{split}
     }
where 
\Equationnoeqn{eq:unitarytr}
{
     \begin{split}
     \gamma_{\vk \sigma}^{(\nu)} 
     = & 
     \sum_{X} [u_{\vk \sigma}^{(X\nu)}]^{*} c_{\vk \sigma}^{(X)} , 
     \\ 
     c_{\vk \sigma}^{(X)} 
     = & 
     \sum_{\nu} u_{\vk \sigma}^{(X\nu)}
     \gamma_{\vk \sigma}^{(\nu)} . 
     \end{split}
     }
For given $m$ and $\mu$, 
the total energy of the system is expressed as 
\Equationnoeqn{eq:Emmu}
{
     \langle H \rangle 
     = \sum_{\nu} {\sum_{\vk, \sigma}}' 
         f(E_{\vk \sigma}^{(\nu)}) 
           E_{\vk \sigma}^{(\nu)} + {\hat E}_0(m,\mu) 
     \equiv {\hat E}(m,\mu) . 
     }
The extremum conditions 
$\partial {\hat E}/\partial m = 0$ 
and 
$\partial {\hat E}/\partial \mu = 0$ 
lead to the self-consistent equations 
\Equationnoeqn{eq:mSCE}
{
     \begin{split}
     m 
     & 
     = \frac{1}{2} 
       \sum_{\sigma} 
       s_{\sigma} 
       \langle c_{i \sigma}^{({\rm A})\dagger}
               c_{i \sigma}^{({\rm A})} \rangle 
     \\ 
     & 
     = \frac{1}{2} 
       \sum_{\sigma} 
       s_{\sigma} 
       \frac{4}{N} {\sum_{\vk}}' 
       [u_{\vk\sigma}^{({\rm A}\nu)}]^{*} 
       f(E_{\vk\sigma}^{(\nu)}) 
       u_{\vk\sigma}^{({\rm A}\nu)}
     \end{split}
     }
and 
\Equation{eq:nSCE}
{
     n 
     = 
       \frac{1}{N} 
       \sum_{i, \sigma} \langle n_{i \sigma} \rangle 
     = 
       \frac{1}{4} 
       \sum_{\nu, \sigma}
       \frac{4}{N} 
       {\sum_{\vk}}' 
       f(E_{\vk\sigma}^{(\nu)}) . 
     }
The resultant total energy $E$ is equal to ${\hat E}(m,\mu)$ 
with self-consistent solutions for $m$ and $\mu$. 
We define 
\Equation{eq:DeltaE}
{
     \Delta E \equiv \frac{1}{N} (E - E_{\rm PM}) , 
     }
where $E_{\rm PM}$ is the energy of the paramagnetic state ($m = 0$).
The contributions of the term $- \mu \sum_{i} (-n)$ in \eq.{eq:H0} 
to $E$ and $E_{\rm PM}$ do not cancel out in $\Delta E$, 
because $\mu$ varies with $m$ for a fixed value of $n$. 
Hence, the term cannot be omitted in the evaluation of $\Delta E$ either.

\section{\label{sec:results}
Results 
}

In this section, we present the numerical results for the system with 
$N = 1024 \times 1024$. 
We have compared the results with those for $N = 2048 \times 2048$ 
and confirmed that the results are practically in the thermodynamic limit. 
We assume $n = 1$ 
considering the application of the theory 
to the $\lambda$-Fe system.

\subsection{\label{sec:lambdaFe}
Application to the $\lambda$-Fe system 
}

\begin{figure}[htbp]
\begin{center}
\begin{tabular}{c}
\includegraphics[width=7.2cm]
{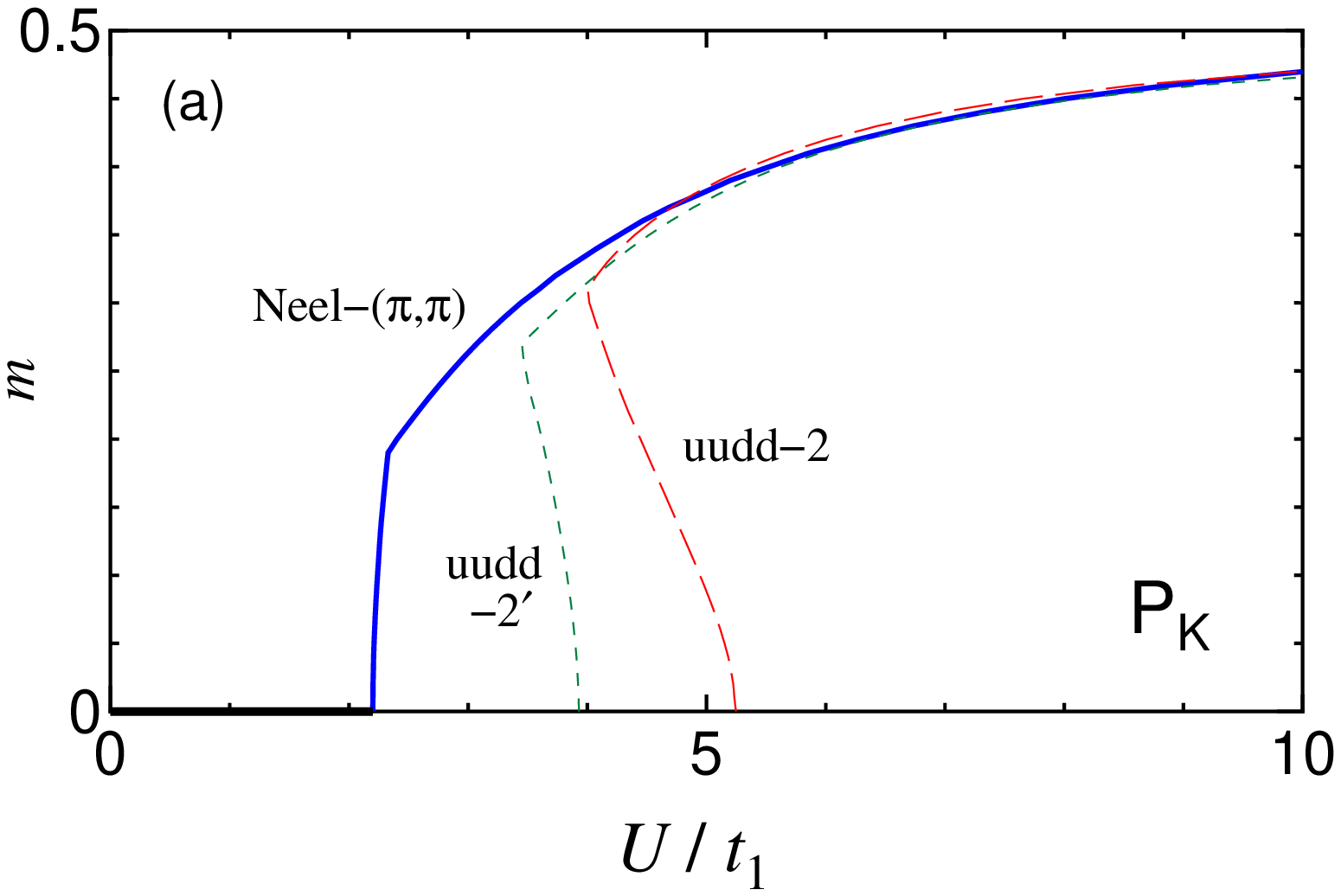}
\\[8pt]
\includegraphics[width=7.2cm]
{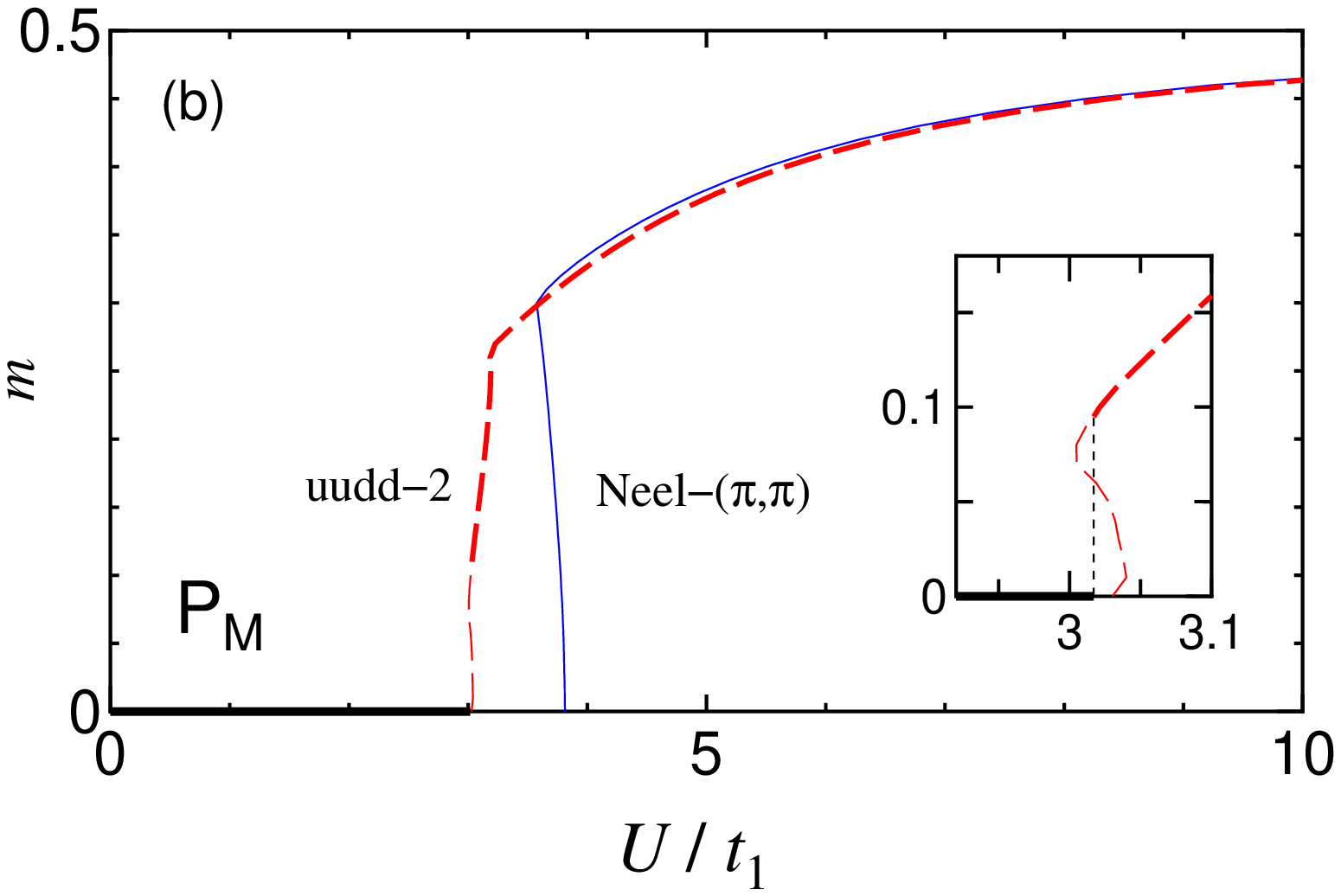}
\\[-8pt]
\end{tabular}
\end{center}
\caption{
(Color online) 
Sublattice magnetizations for parameter sets 
(a) {\PK} and (b) {\PM}. 
The blue solid, red dashed, and green short-dashed curves 
are the sublattice magnetizations $m$ 
for the {\Neelpipi}, uudd-2, and uudd-$2'$ states, respectively. 
The thick and thin curves show the solutions with the lowest energy 
and 
those of the states with relatively higher energies, respectively. 
The black thick solid line of $m = 0$ 
shows the paramagnetic state. 
In the inset of the lower panel (b), 
the vertical black thin dotted lines show 
the jump in $m$ at $U = U_{\rm c}$ 
between the uudd-2 and paramagnetic states. 
} 
\label{fig:mpiPKPM} 
\end{figure}

\begin{figure}[htbp]
\begin{center}
\begin{tabular}{c}
\\
\multicolumn{1}{l}{(a)} 
\\[-8pt] 
\includegraphics[width=7.2cm]
{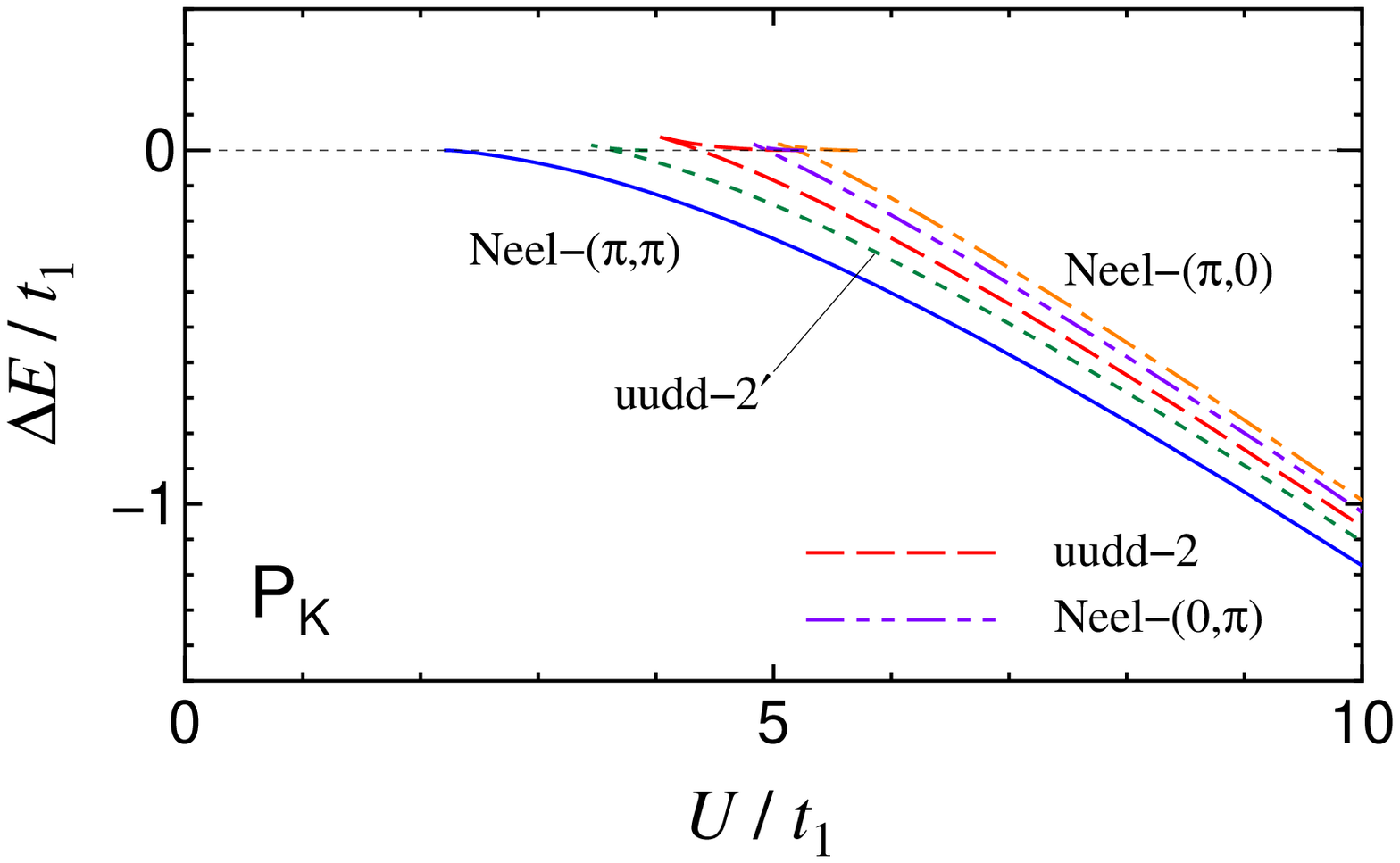}
\\[8pt]
\\
\multicolumn{1}{l}{(b)} 
\\[-8pt] 
\includegraphics[width=7.2cm]
{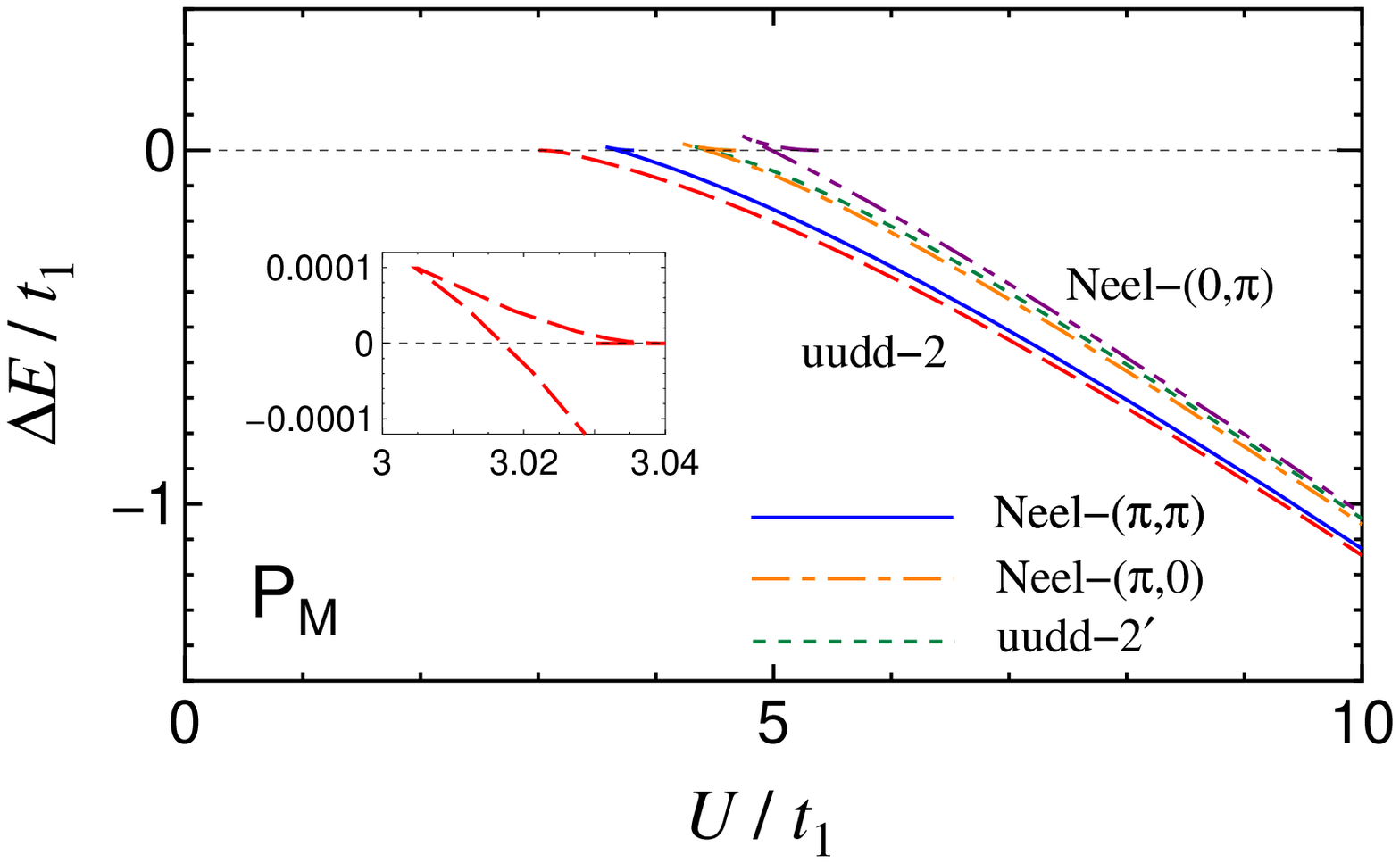}
\\[-8pt]
\end{tabular}
\end{center}
\caption{
(Color online) 
Total energies of the five collinear phases 
near the quantum critical point 
for parameter sets (a) {\PK} and (b) {\PM}. 
The blue solid, 
red dashed, 
green short-dashed, 
orange dot-dashed, 
and 
purple two-dot-dashed curves 
show $\Delta E$ of 
the {\Neelpipi}, uudd-2, uudd-$2'$, {\Neelpizero}, 
and {\Neelzeropi} states, respectively. 
} 
\label{fig:dE_U_PKPM} 
\end{figure}

Figure~\ref{fig:mpiPKPM} shows the behaviors of the solutions for 
the sublattice magnetizations $m$ as functions of $U$ 
for parameter sets {\PK} and {\PM} shown in Table~\ref{table:PMPK}, 
which are candidates for the parameter set of $\lambda$-Fe. 
The value of $U$ is unknown, but it is possibly close to 
$U_{\rm c}$.~\cite{NoteQCP} 
The solutions for $m$ of the states 
having higher energies are omitted in Fig.~\ref{fig:mpiPKPM} 
(see Fig.~\ref{fig:dE_U_PKPM}). 
At the resulting $U_{\rm c}$, the solution for the {\Neelpipi} state 
for {\PK} exhibits the second-order transition 
[Fig.~\ref{fig:mpiPKPM}(a)], 
whereas that for the uudd-2 state for {\PM} exhibits 
the first-order transition; i.e., 
the value of $m$ jumps between the uudd-2 and paramagnetic states 
[Fig.~\ref{fig:mpiPKPM}(b)].

The total energies ${\hat E}(m,\mu)$ are calculated 
for the five collinear states 
with the solutions for $m$ and $\mu$. 
Figure~\ref{fig:dE_U_PKPM} plots $\Delta E$ for the five collinear states 
near the quantum critical point. 
For parameter set {\PK}, 
the {\Neelpipi} phase has the lowest energy, 
whereas for parameter set {\PM}, 
the uudd-2 phase has the lowest energy, 
slightly below the {\Neelpipi} phase with the second-lowest energy. 
Hence, the ground state of the $\lambda$-Fe system 
is most likely the {\Neelpipi} phase or the uudd-2 phase. 
We find again that for ${\rm P}_{\rm K}$, 
the transition between the {\Neelpipi} and paramagnetic states 
is of the second order, 
whereas for ${\rm P}_{\rm M}$, 
the transition between the uudd-2 and paramagnetic states 
is of the first order; 
however, 
the jump in the energy slope between these states 
is extremely small, 
as shown in the inset of Fig.~\ref{fig:dE_U_PKPM}(b).

\subsection{\label{sec:extend}
Extending parameter region 
}

In this subsection, 
we extend the range of the parameters.

\subsubsection{\label{sec:DeltaE-U_with_t3}
Effect of the shift in $t_3$ and $\rimb$ 
on $\Delta E(U)$
}

\begin{figure}[htbp]
\begin{center}
\begin{tabular}{c}
\\
\multicolumn{1}{l}{(a)} 
\\
\includegraphics[width=7.2cm]
{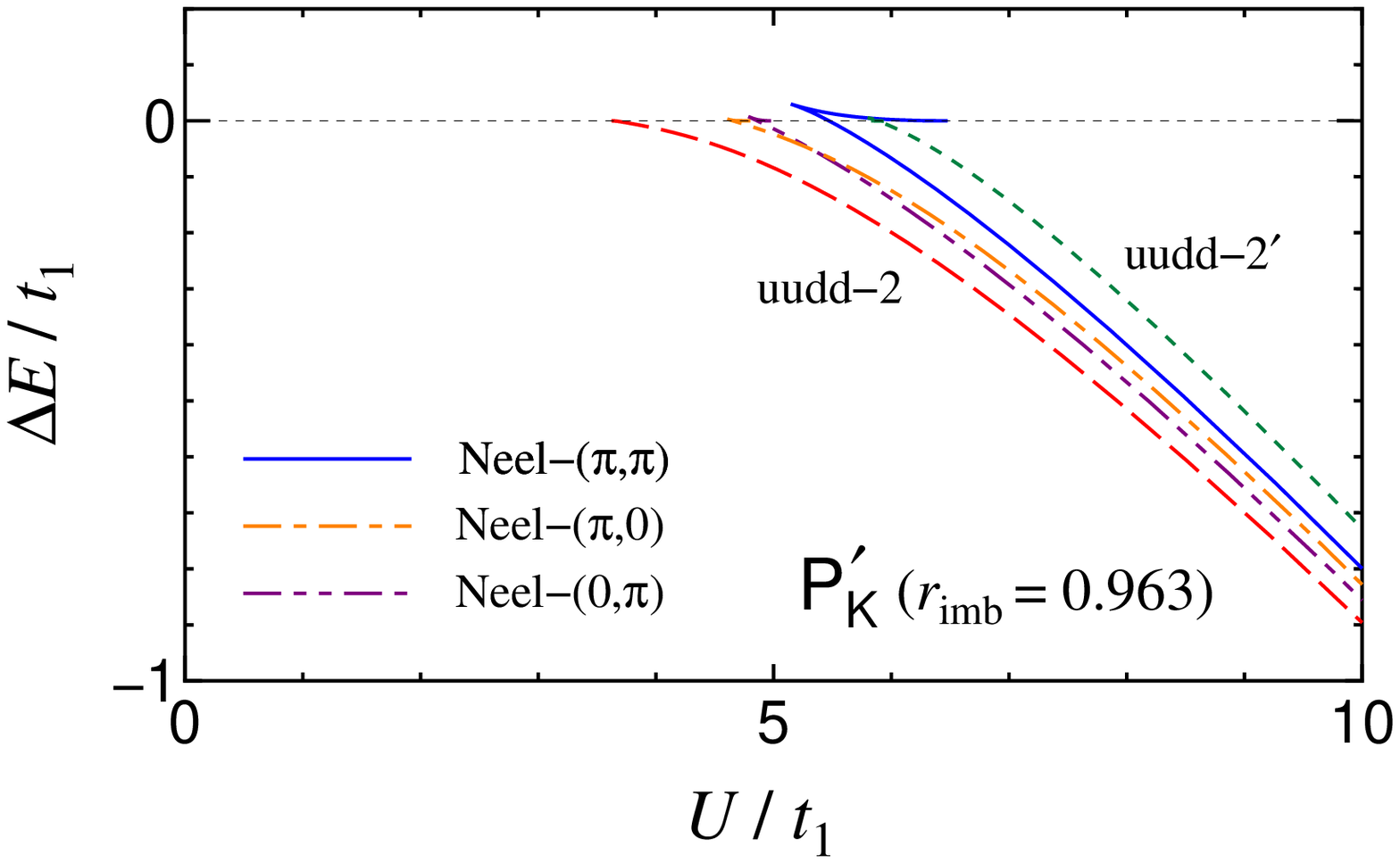} 
\\
\multicolumn{1}{l}{(b)} 
\\
\includegraphics[width=7.2cm]
{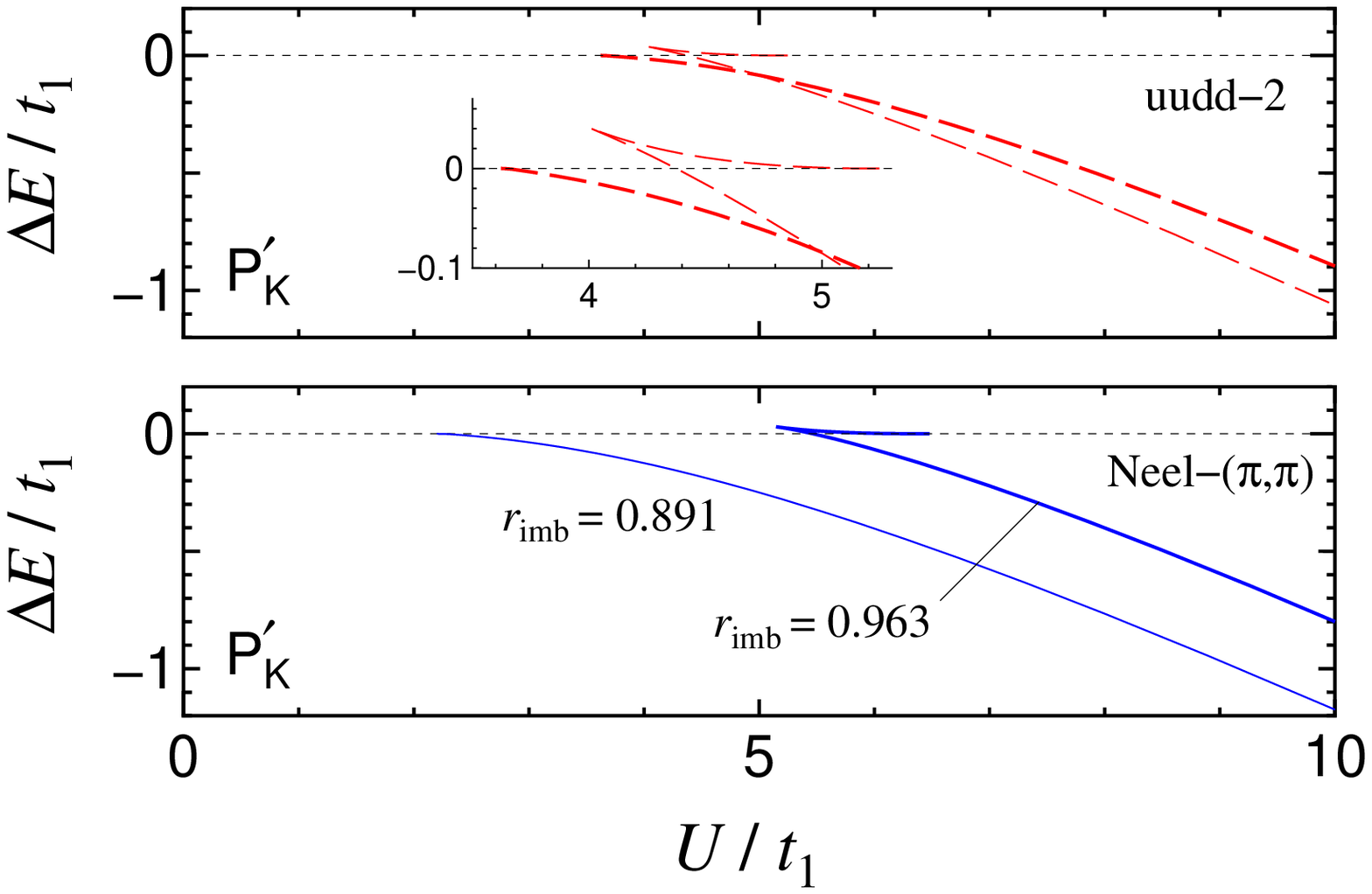}
\\[-8pt]
\end{tabular}
\end{center}
\caption{
(Color online) 
Total energies 
for parameter set {\PKprim} with $t_3 = 7.5 \times 10^{-2}~{\rm eV}$, 
which leads to $\rimb = 0.963$. 
(a) The energies of the five collinear phases. 
(b) Comparison with the results for {\PK}, 
which are represented by the thin curves. 
For {\PK}, $\rimb = 0.891$. 
The legend is the same as that in Fig.~\ref{fig:dE_U_PKPM}. 
}
\label{fig:dE_U_PK_increaset3} 
\end{figure}

Figure~\ref{fig:dE_U_PK_increaset3} shows the value of 
$\Delta E(U)$ for {\PKprim}, 
in which $t_3 = 7.5 \times 10^{-2}~{\rm eV}$. 
The critical value $U_{\rm c}$ for each magnetic structure 
is given by $\Delta E(U_{\rm c}) = 0$. 
The increase in $t_3$ results in an increase in $\rimb$, 
i.e., 
the increase of the imbalance of the anisotropies. 
In contrast to {\PK}, 
for which the {\Neelpipi} state is the ground state, 
the uudd-2 state becomes the ground state for {\PKprim} 
as shown in Fig.~\ref{fig:dE_U_PK_increaset3}(a). 
This figure also shows that the transition at $U_{\rm c}$ 
is of the second order. 
When $\rimb$ increases, 
the energy of the {\Neelpipi} state increases as shown in 
Fig.~\ref{fig:dE_U_PK_increaset3}(b), 
whereas that of the uudd-2 state decreases 
in the close vicinity of $U_{\rm c}$; hence, 
$U_{\rm c}$ of the uudd-2 phase is slightly enhanced.

As a physical interpretation of 
the effect of the spatial-anisotropy imbalance, 
when the imbalance is large, 
the two types of bond triangles tend to have 
different frustrated spin structures; 
hence, 
the uudd states shown in Fig.~\ref{fig:uudd} 
are favored more than 
the {\Neel} states shown in Fig.~\ref{fig:Neel}. 
In particular, as $t_3$ increases, 
the exchange coupling constant $J_3 \sim t_3^2/U$ increases, 
favoring an antiparallel configuration 
of the two spins on the sites connected by bond 3; 
hence, the uudd-2 state is more stabilized than 
the uudd-$2'$ state, when $t_3$ is large.

\begin{figure}[htbp]
\begin{center}
\begin{tabular}{c}
\\
\multicolumn{1}{l}{(a)} 
\\[-8pt] 
\includegraphics[width=7.2cm]
{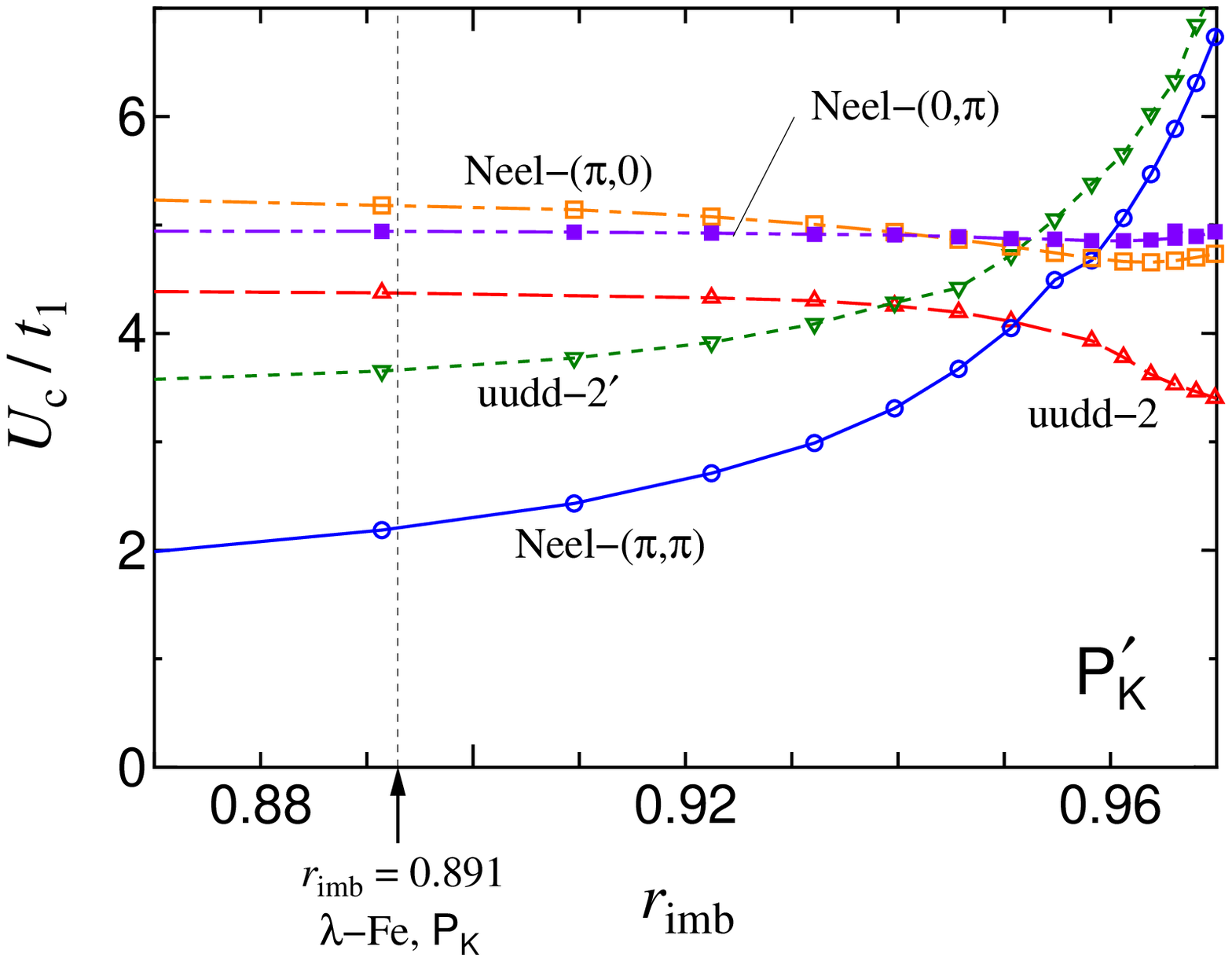} \\ 
\\[8pt]
\\
\multicolumn{1}{l}{(b)} 
\\[-8pt] 
\includegraphics[width=7.2cm]
{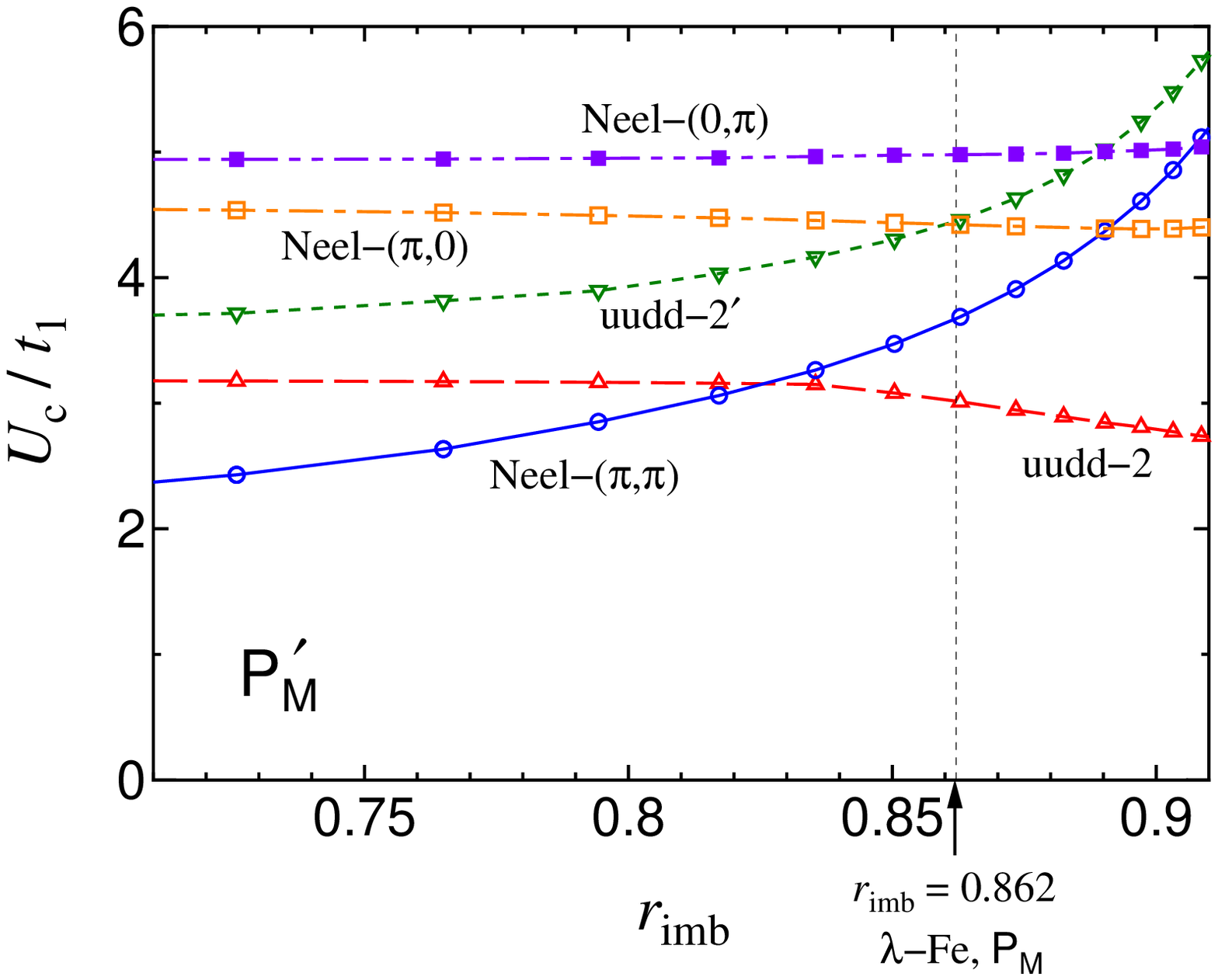} \\ 
\\[-8pt]
\end{tabular}
\end{center}
\caption{
(Color online) 
$U_{\rm c}(\rimb)$ of the five collinear states for parameter sets 
(a) {\PKprim} and (b) {\PMprim}, 
in which $t_3$ is variable. 
The open circles, open squares, closed squares, open triangles, 
and open inverted triangles are 
the results for the {\Neelpipi}, {\Neelpizero}, {\Neelzeropi}, 
uudd-2, and uudd-$2'$ states, respectively. 
The curves are guides to the eye. 
} 
\label{fig:Uc_r} 
\end{figure}

\begin{figure}[htbp]
\begin{center}
\begin{tabular}{c}
\\
\multicolumn{1}{l}{(a)} 
\\[-8pt] 
\includegraphics[width=7.2cm]
{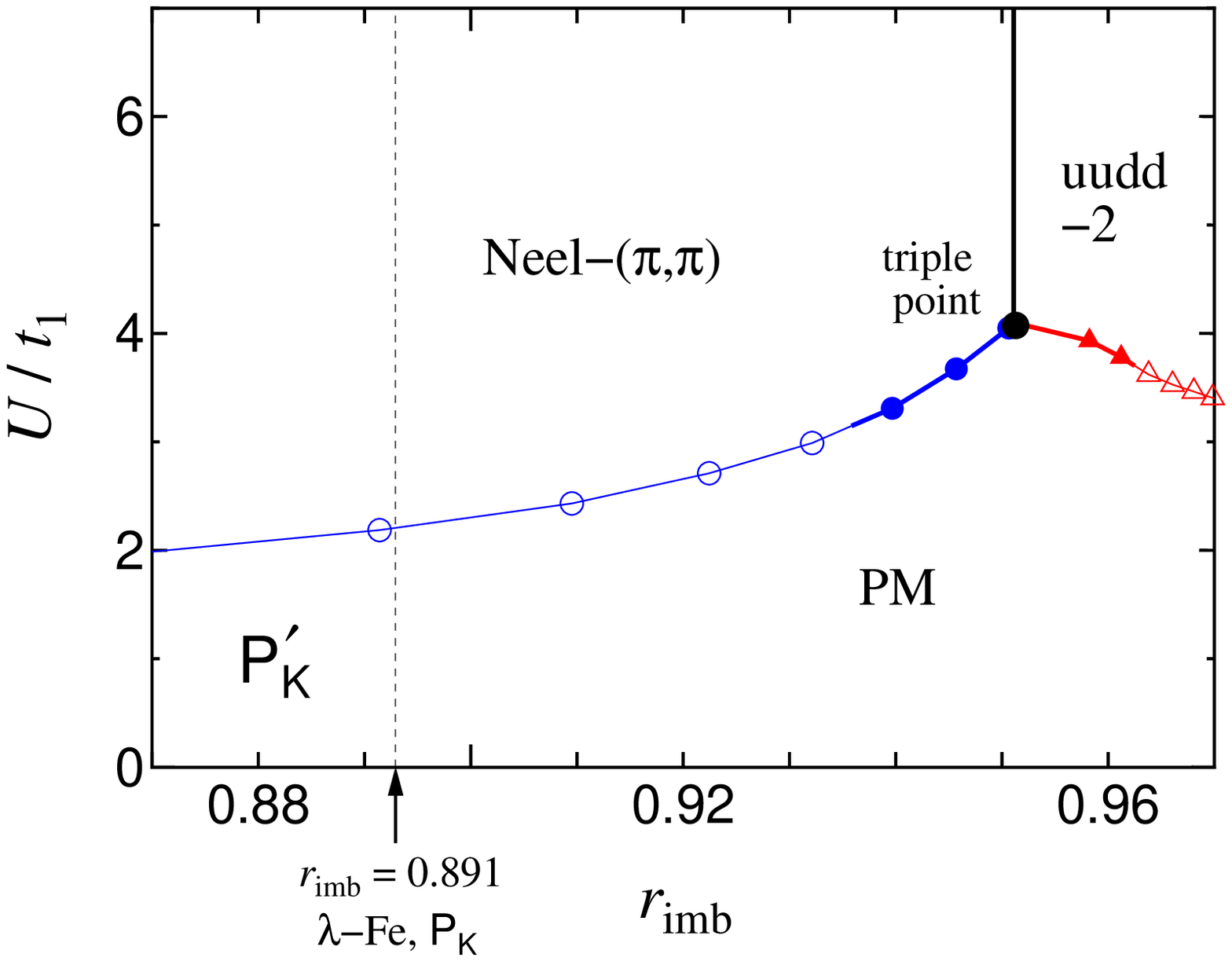}
\\[8pt]
\\
\multicolumn{1}{l}{(b)} 
\\[-8pt] 
\includegraphics[width=7.2cm]
{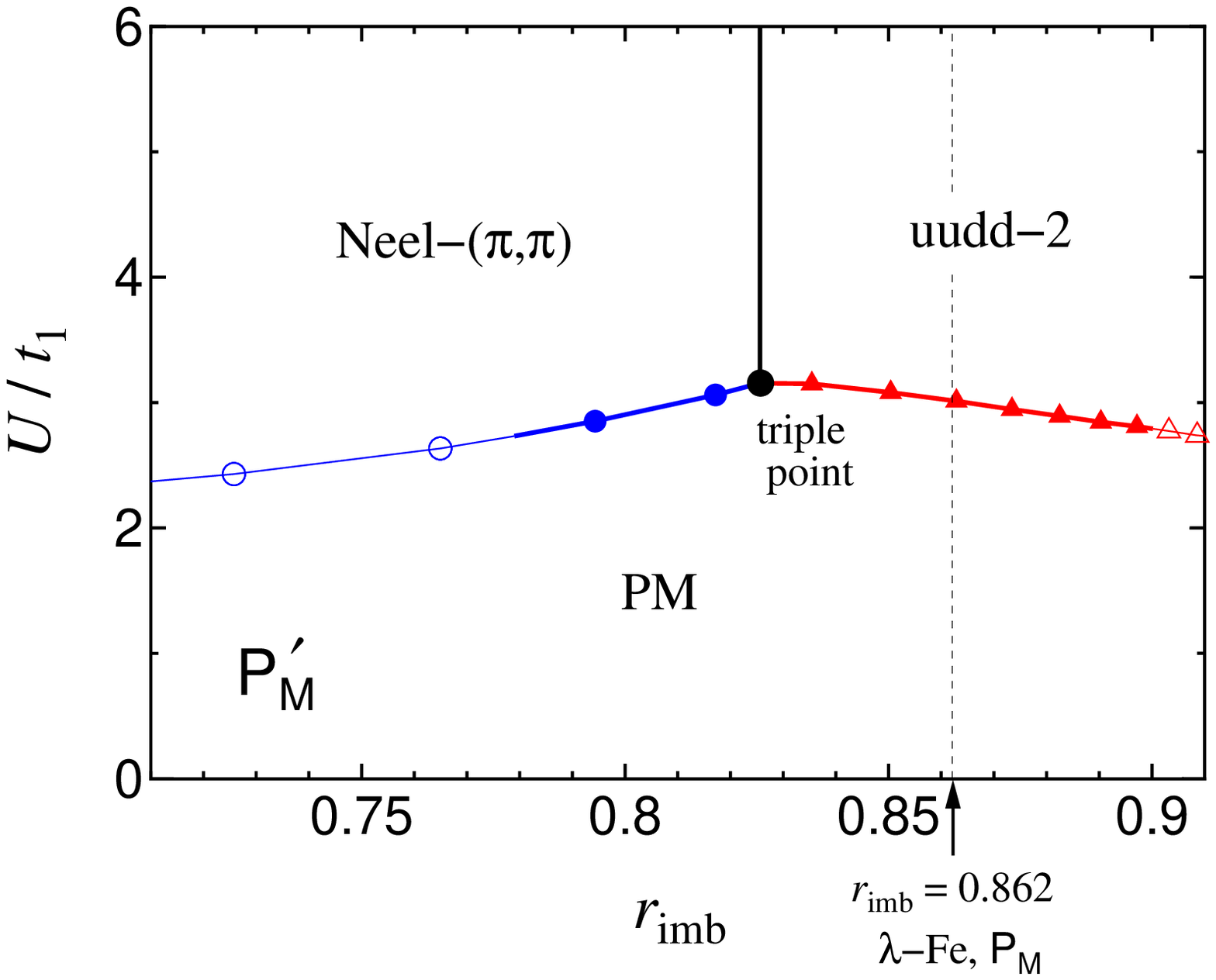}
\\[-8pt]
\end{tabular}
\end{center}
\caption{
(Color online) 
Phase diagrams of the systems with 
parameter sets {\PKprim} and {\PMprim}. 
The blue circles and red triangles are the points on the phase boundary 
of the {\Neelpipi} phase 
and those between the uudd-2 and paramagnetic (PM) phases, respectively. 
The closed and open symbols indicate first-order 
and second-order transitions, 
respectively. 
The black thick solid lines show first-order transition 
lines between 
the {\Neelpipi} and uudd-2 states. 
The blue and red curves are guides to the eye, 
where the thick and thin curves correspond 
to first- and second-order transitions, 
respectively. 
The large black closed circles show the triple points. 
} 
\label{fig:PhaseD} 
\end{figure}

\subsubsection{\label{sec:Uc_rimb_dep}
$U_{\rm c}(\rimb)$ and phase diagrams 
}

Using parameter sets {\PKprim} and {\PMprim}, 
in which $t_3$ is variable, 
we examine the influence of the shift in $\rimb$. 
Figure~\ref{fig:Uc_r} shows the behaviors of 
the critical values $U_{\rm c}$ of the five collinear states, 
below which their energies are higher than the energy of 
the paramagnetic state. 
The {\Neelpipi} and uudd-2 states are suppressed and enhanced 
by the increase in $\rimb$, respectively, 
which implies that the imbalance of the spatial anisotropies 
favors the uudd-2 state. 
The other collinear states, such as 
the {\Neelpizero}, {\Neelzeropi}, and uudd-$2'$ states, 
have higher energies in the present parameter ranges.

The phase diagrams in the $U$-$\rimb$ plane are shown in 
Fig.~\ref{fig:PhaseD}. 
For both parameter sets {\PKprim} and {\PMprim}, 
they contain areas of the {\Neelpipi}, 
uudd-2, and paramagnetic phases, 
and the uudd-2 phase occupies an area in which 
$\rimb$ is large. 
The other collinear states do not appear 
in the parameter ranges of these figures. 
In both phase diagrams, a triple point 
of the two antiferromagnetic phases and 
the paramagnetic phase exists, 
near which all the transitions are of the first order.
The phase boundary between the {\Neelpipi} and uudd-2 states 
is parallel to the $U$-axis, 
and the transition is of the first order everywhere on the boundary. 
By contrast, the transitions from the two antiferromagnetic phases to 
the paramagnetic phase are of the second order 
when the system is far from the triple point. 
This feature is common to both {\PKprim} and {\PMprim}.

The values of $\rimb$ for the $\lambda$-Fe system are 
shown by the vertical thin dotted lines 
in Fig.~\ref{fig:PhaseD}. 
The increase in $U$ does not change the magnetic structure, 
provided that an antiferromagnetic state is the ground state 
within the present parameter ranges. 
By contrast, small changes in $\rimb$ 
($\Delta \rimb \gsim 7\%$ for {\PK} 
and $|\Delta \rimb| \gsim 4\%$ for {\PM}) 
can change the magnetic structure.

\section{\label{sec:summary and discussion} 
Summary and Discussion 
}

We examined the magnetic structure 
of electron systems on the ESCATL 
while focusing on the collinear states 
at $T = 0$, 
and we adopted the Hubbard model and mean-field approximation. 
Within the parameter range, 
the {\Neelpipi} state and a uudd state occur, and 
the transitions to these states can be of the first or second order 
depending on the parameters applied. 
The uudd states are favored under a large imbalance of 
the spatial anisotropies, 
which is a unique feature of the ESCATL, 
whereas the {\Neelpipi} state occurs in wider parameter ranges. 
In the application to the $\lambda$-Fe system, 
we used candidate parameter sets {\PK} and {\PM}.~\cite{NoteHot00} 
As a result, it was found that for {\PK}, 
a second-order transition to the {\Neelpipi} state occurs at $U_{\rm c}$, 
whereas for {\PM}, a first-order transition to 
the uudd-2 phase occurs at $U_{\rm c}$.

By extending parameter regions, 
we found triple points of the {\Neelpipi}, 
uudd-2, and paramagnetic phases, 
near which all transitions are of the first order. 
The transitions from the {\Neelpipi} and uudd-2 phases 
to the paramagnetic phase are of the second order unless 
the system is close to the triple points both for {\PKprim} and {\PMprim}. 
It should be examined in future research whether 
the existence of a triple point and the changes 
of the order of the transitions are 
universal features in the electron systems on the ESCATL.

For the $\lambda$-Fe system, 
the nesting vector $(\pi/c,0)$ obtained by previous studies 
suggests that the antiferromagnetic state has 
the modulation vector $(\pi/c,0)$ or $(\pi/c,\pi/a)$ 
as explained in Sect.~\ref{sec:introduction}. 
Between these states, 
the present result supports the antiferromagnetic state with $(\pi/c,\pi/a)$. 
In addition, 
the present study showed that 
the uudd phase can be the ground state of this system 
depending on the parameter values. 
The results are consistent with those in the classical spin system.~\cite{Sak17}

In a previous study, 
the magnetic structure of the $\lambda$-Fe system was 
examined within a similar mean-field theory.~\cite{Hot00} 
However, their model differs from our proposed 
model in many ways, 
resulting in discrepancies in the results. 
One of the significant differences between the two models lies in the lattice structures. 
In the previous model, the lattice site in the $\pi$-electron system 
corresponds to each BETS molecule, 
whereas in the present model, each dimer of the molecules 
is regarded as a lattice site. 
This results in a difference in the physical meaning of 
the on-site ``$U$.'' 
For example, 
$U$ in the previous model works only when the two electrons 
(or holes) are on the same BETS molecule, 
whereas the on-site $U$ in the present dimer model 
works when they are in the same dimer. 
Hence, the present $U$ includes an effect 
of the Coulomb repulsion between two electrons 
on different BETS molecules in the same dimer; however, 
such an interaction is ignored in the previous study. 
Thus, the effect of ``$U$'' must differ between the two models. 
For example, in the previous model, 
when $U$ increases, the system undergoes successive transitions, 
which were not found in the present study. 
The difference in the lattice structure also results in 
a difference in the filling of the relevant band having Fermi surfaces: 
In the present model, it is half-filled, 
which favors the insulating phase as observed in 
the $\lambda$-Fe system at low temperatures, 
whereas it is quarter-filled in the previous model. 
Meanwhile, 
the magnetic structure inside the dimer considered in the previous study 
is beyond the scope of this paper. 
The number of the sublattices in the mean-field approximation 
is different between the two theories. 
Another important difference is the presence of 
the 3d-spins: 
the previous model contains both $\pi$-electrons and 3d-spins, 
whereas we examined a pure $\pi$-electron system without d-spins 
on the basis of the current knowledge 
that the $\pi$-electron system is the principal component 
and the d-spins are passive in the exchange field created 
by the $\pi$-electrons.~\cite{Aki09}

Although the phase diagrams do not contain the areas of 
the {\Neelpizero} and {\Neelzeropi} states 
in the parameter regions examined in the present study, 
these states must occur depending on the parameters 
in the present itinerant system 
as well as in the classical localized spin system. 
Furthermore, although we neglected the spiral spin state, 
according to the experimental results~\cite{Osh17,Sas01,Tok05,Aki11,Sat98,Tok97,Ito16} 
for the $\lambda$-Fe system, 
it may occur in other compounds 
including those that are yet undiscovered. 
A search for richer phase diagrams is 
a promising future research direction.

In the $\lambda$-Fe system, 
the magnetic long-range order 
is considered to be stabilized 
by the factors that originate from the d-spins 
of the ${\rm FeCl_4}$ anions,~\cite{Shi14,Shi17,Shi18} 
such as the anisotropy in the spin space 
and/or the enhanced three dimensionality. 
As mentioned in Sect.~\ref{sec:introduction}, 
the present mean-field approximation implicitly assumes such factors. 
In future studies, 
improved theories beyond the mean-field approximation 
must explicitly incorporate these factors 
so that the stable magnetic 
long-range order in the $\lambda$-Fe system 
is reproduced.

It can be expected that the above results concerning the energies of 
the antiferromagnetic states 
are hardly affected by the d-spins, 
which are not incorporated in the present model, 
because the interactions in the $\pi$-electron system are much stronger 
than the other interactions 
(those in the d-spin system and 
those between the $\pi$-electrons and d-spins).

In conclusion, 
itinerant electron systems on the ESCATL 
as well as the localized spin system 
can exhibit uudd phases 
when the imbalance of the anisotropies is large. 
Within the parameter range examined, 
the transitions at $U_{\rm c}$ 
between the paramagnetic phase 
and the antiferromagnetic phase, 
such as the {\Neelpipi} and uudd phases, 
are of the second order, 
except for a small parameter region 
near the junction of the three phase boundaries. 
Near the junction, 
all transitions are of the first order; 
hence, it is a triple point. 
The ground state of the $\lambda$-Fe system 
is most likely the {\Neelpipi} state or a uudd state.

\mbox{}


\begin{acknowledgments}
The authors would like to thank 
\mbox{Yutaka Nishio}, 
Sinya Uji, 
Yugo Oshima, 
Takaaki Minamidate, 
Shuhei Fukuoka, 
and 
\mbox{Takuya Kobayashi} 
for useful discussions. 
\end{acknowledgments}

\mbox{}

\appendix
\section{
Explicit forms of the matrices ${\hat {\cal E}}_{\vk\sigma}$ 
}
\label{app:ExplicitE}

This appendix shows the matrix elements of 
\Equationnoeqn{eq:hatcalE_elements} 
{
     {\hat {\cal E}}_{\vk\sigma} 
     = \left ( 
         \begin{array}{cccc}
         {\tilde \xi}_{\vk\sigma}^{\, \rm (AA)}
         & 
         {\tilde \xi}_{\vk}^{\, \rm (AA')}
         & 
         {\tilde \xi}_{\vk}^{\, \rm (AB)}
         & 
         {\tilde \xi}_{\vk}^{\, \rm (AB')}
         \\[4pt]
         {\tilde \xi}_{\vk}^{\, \rm (A'A)}
         & 
         {\tilde \xi}_{\vk\sigma}^{\, \rm (A'A')}
         & 
         {\tilde \xi}_{\vk}^{\, \rm (A'B)}
         & 
         {\tilde \xi}_{\vk}^{\, \rm (A'B')}
         \\[4pt]             
         {\tilde \xi}_{\vk}^{\, \rm (BA)}
         &                   
         {\tilde \xi}_{\vk}^{\, \rm (BA')}
         &                   
         {\tilde \xi}_{\vk\sigma}^{\, \rm (BB)}
         &                   
         {\tilde \xi}_{\vk}^{\, \rm (BB')}
         \\[4pt]             
         {\tilde \xi}_{\vk}^{\, \rm (B'A)}
         &                   
         {\tilde \xi}_{\vk}^{\, \rm (B'A')}
         & 
         {\tilde \xi}_{\vk}^{\, \rm (B'B)}
         & 
         {\tilde \xi}_{\vk\sigma}^{\, \rm (B'B')}
         \end{array}
       \right ) . 
     }
The diagonal elements are common to 
all the five collinear states 
and expressed as 
$$
     {\tilde \xi}_{\vk}^{(XX)} = - s_{X} s_{\sigma} U m - \mu , 
     $$
where $X = {\rm A}, {\rm A'}, {\rm B}$, and ${\rm B'}$. 
The off-diagonal elements 
${\tilde \xi}_{\vk}^{(XY)} = [{\tilde \xi}_{\vk}^{(YX)}]^*$ 
are expressed as follows. 
For the {\Neelpipi} state, 
\Equationnoeqn{eq:Neelpipi}
{
     \begin{split}
     {\tilde \xi}_{\vk}^{\, \rm (AB)} 
     & =   
     {\tilde \xi}_{\vk}^{\, \rm (A'B')} 
     = 2 t_1 \cos k_x , 
     \\ 
     {\tilde \xi}_{\vk}^{\, \rm (AB')} 
     & =   
     {\tilde \xi}_{\vk}^{\, \rm (A'B)} 
     =   
     t_2 e^{- i k_y} + t_2' e^{ i k_y} , 
     \\ 
     {\tilde \xi}_{\vk}^{\, \rm (AA')} 
     & =   
     {\tilde \xi}_{\vk}^{\, \rm (BB')} 
     =   
     t_3 e^{- i (k_x + k_y)} + t_3' e^{ i (k_x + k_y)} . 
     \end{split} 
     }
For the {\Neelpizero} state, 
\Equationnoeqn{eq:Neelpizero}
{
     \begin{split}
     {\tilde \xi}_{\vk}^{\, \rm (AB)} 
     & =   
     {\tilde \xi}_{\vk}^{\, \rm (A'B')} 
     = 2 t_1 \cos k_x , 
     \\ 
     {\tilde \xi}_{\vk}^{\, \rm (AA')} 
     & =   
     {\tilde \xi}_{\vk}^{\, \rm (BB')} 
     =   
     t_2 e^{- i k_y} + t_2' e^{ i k_y} , 
     \\ 
     {\tilde \xi}_{\vk}^{\, \rm (AB')} 
     & =   
     {\tilde \xi}_{\vk}^{\, \rm (BA')} 
     =   
     t_3 e^{- i (k_x + k_y)} + t_3' e^{ i (k_x + k_y)} . 
     \\ 
     \end{split} 
     }
For the {\Neelzeropi} state, 
\Equationnoeqn{eq:Neelzeropi}
{
     \begin{split}
     {\tilde \xi}_{\vk}^{\, \rm (AA')} 
     & =   
     {\tilde \xi}_{\vk}^{\, \rm (BB')} 
     = 2 t_1 \cos k_x , 
     \\ 
     {\tilde \xi}_{\vk}^{\, \rm (AB)} 
     & =   
     {\tilde \xi}_{\vk}^{\, \rm (A'B')} 
     =   
     t_2 e^{- i k_y} + t_2' e^{ i k_y} , 
     \\ 
     {\tilde \xi}_{\vk}^{\, \rm (AB')} 
     & =   
     {\tilde \xi}_{\vk}^{\, \rm (BA')} 
     =   
     t_3 e^{- i (k_x + k_y)} + t_3' e^{ i (k_x + k_y)} . 
     \\ 
     \end{split} 
     }
For the uudd-2 state, 
\Equationnoeqn{eq:uudd2}
{
     \begin{split}
     {\tilde \xi}_{\vk}^{\, \rm (AB)} 
     & =   
     {\tilde \xi}_{\vk}^{\, \rm (A'B')} 
     = 2 t_1 \cos k_x , 
     \\ 
     {\tilde \xi}_{\vk}^{\, \rm (AA')} 
     & =   
     {\tilde \xi}_{\vk}^{\, \rm (BB')} 
     =   
     t_2 e^{- i k_y} + t_3' e^{ i (k_x + k_y)} , 
     \\ 
     {\tilde \xi}_{\vk}^{\, \rm (A'B)} 
     & =   
     {\tilde \xi}_{\vk}^{\, \rm (B'A)} 
     =   
     t_2' e^{i k_y} + t_3 e^{ - i (k_x + k_y)} . 
     \\ 
     \end{split} 
     }
For the uudd-$2'$ state, 
\Equationnoeqn{eq:uudd2prim}
{
     \begin{split}
     {\tilde \xi}_{\vk}^{\, \rm (AB)} 
     & =   
     {\tilde \xi}_{\vk}^{\, \rm (A'B')} 
     = 2 t_1 \cos k_x , 
     \\ 
     {\tilde \xi}_{\vk}^{\, \rm (A'B)} 
     & =   
     {\tilde \xi}_{\vk}^{\, \rm (B'A)} 
     =   
     t_2 e^{- i k_y} + t_3' e^{ i (k_x + k_y)} , 
     \\ 
     {\tilde \xi}_{\vk}^{\, \rm (BB')} 
     & =   
     {\tilde \xi}_{\vk}^{\, \rm (AA')} 
     =   
     t_2' e^{- i k_y} + t_3 e^{ i (k_x + k_y)} . 
     \\ 
     \end{split} 
     }




\begin{thebibliography}{99}
\bibitem{Notepid}
  This fact was suggested by 
  the results of the specific heat measurements.~\cite{Aki09,Aki11}
\bibitem{NoteHot00}
  The values of the transfer integrals obtained by 
  Kobayashi and Mori 
  are presented in Ref.~\citen{Hot00}. 
  Their results do not coincide; 
  however, 
  because explicit physical conditions assumed for their calculations 
  are not presented, 
  the reason for the difference is unknown. 
  We adopt them as candidates for the parameters 
  of the $\lambda$-Fe system.
\bibitem{NoteQCP} 
  An argument in Ref.~\citen{Shi17} suggests that 
  the $\lambda$-Fe system 
  is close to the quantum critical point 
  ($U_{\rm c} - U \ll U$). 
  A simple analysis leads to the expression of 
  the antiferromagnetic transition temperature $\TAF(x)$ 
  in the mixed crystal $\lambda$-${\rm Fe}_x{\rm Ga}_{1-x}$: 
  $$
       \TAF(x) 
         = x \frac{S(S+1)}{6} \frac{J_{\pi {\rm d}}^2}{U_{\rm c} - U}, 
       $$
  where $S = 5/2$ is the magnitude of the d-spin, 
  and $J_{\pi {\rm d}}$ is the exchange coupling constant 
  between the $\pi$-electrons and the d-spins. 
  This expression explains the experimental observations 
  that $\TAF(x) \propto x$. 
  From the experimental observation that $\TAF(1) \sim J_{\pi {\rm d}}$, 
  it follows that $U_{\rm c} - U \sim J_{\pi {\rm d}} \sim \TAF(1) \ll U$.
\bibitem{Notemun}
  The term $- \mu \sum_{i} (-n)$ in \eq.{eq:H0} 
  cannot be eliminated 
  by an appropriate choice of the origin of the energy; 
  omitting the term in the evaluation of $E$ leads to unphysical results. 
  See the argument below \eq.{eq:DeltaE}.
\end{thebibliography}
\end{document}